# The Chaplygin gas as a model for modified teleparallel gravity?

Shambel Sahlu[1,2,a], Joseph Ntahompagaze[1,3], Maye Elmardi[4], Amare Abebe[1,5]

[1] Astronomy and Astrophysics Department, Entoto Observatory and Research Center, Ethiopian Space Science and Technology Institute, Addis Ababa, Ethiopia
[2] Department of Physics, College of Natural and Computational Science, Wolkite University, Wolkite, Ethiopia
[3] Department of Physics, College of Science and Technology, University of Rwanda, Kigali, Rwanda
[4] Center for Space Research, North-West University, Potchefstroom, South Africa
[5] Center for Space Research, North-West University, Mahikeng, South Africa

**Abstract** This paper explores the possibility of treating the exotic Chaplygin-gas (CG) fluid model as some manifestation of an $f(T)$ gravitation. To this end, we use the different cosmological CG equations of state, compare them with the equation of state for the modified teleparallel gravity and reconstruct the corresponding Lagrangian densities. We then explicitly derive the equation of state parameter of the torsion fluid $w_T$ and study its evolution for vacuum-torsion, radiation-torsion, dust-torsion, stiff fluid-torsion and radiation-dust-torsion multi-fluid systems. The obtained Lagrangians have, in general, matter dependence due to the matter-torsion coupling appearing in the energy density and pressure terms of the modified teleparallel gravity theory. For the simplest CG models, however, it is possibly to reconstruct $f(T)$ Lagrangians that depend explicitly on the torsion scalar $T$ only. The preliminary results show that, in addition to providing Chaplygin-gas-like solutions to the modified teleparallel gravitation, which naturally behave like dark matter and dark energy at early and late times respectively, the technique can be used to overcome some of the challenges attributed to the CG cosmological alternative.

## 1 Introduction

The discovery of the accelerated expansion of the Universe has posed one of the biggest challenges for observational and theoretical cosmology and, two decades on, remains an open problem. Several suggestions have been put forward as the possible causes behind this cosmic acceleration, including the dark energy (DE) hypothesis and modified gravity theories. Dark energy is a hypothetical form of energy with negative pressure, acting against gravity, and believed to permeate all of space, accounting for about 68% of the entire energy content of the Universe [1]. Among the most well known natural candidates for dark energy are: a cosmological constant [2], a self-interacting scalar field [3] and the cosmological CG as a non-interacting fluid [4]. In the modified gravity approach, several theories have been constructed by modifying the Einstein–Hilbert action such as $f(T)$ - $T$ being the torsion scalar, $f(R)$ - $R$ being the Ricci scalar, a combination of the two named $f(R, T)$; $f(G)$ - $G$ being Gauss-Bonnet term, etc. Among the earliest such attempts was by Einstein himself, when he used the teleparallel gravity (TG) theory [5] to unify the theory of electromagnetism with gravity theory [6].

In the general relativity (GR) approach, $T$ is assumed to vanish and in TG theory, $R$ is assumed to vanish [7]. Fortunately, the two basic theories of gravity describe the gravitational interaction equivalently. So, torsion is an alternative direction of describing the gravitational field interaction. The energy-momentum tensor (EMT) is the source of curvature in GR and torsion in TG theory [6–15]. More recently, generalizations to the TG theory have been introduced in the form of $f(T)$ gravity, where the action is now a generic function of $T$, rather than T itself. It is a second-order modified gravity theory and understanding its cosmological implications is an active area of research [6,16–21].

The Chaplygin gas (CG), first introduced by Chaplygin [22], had a non-cosmological origin but has recently gained new attention in cosmology due to its negative pressure. Together with a Friedman–Lemaître–Robertson–Walker (FLRW) background in the GR framework, this model can explain the cosmic expansion history for a Universe filled with an exotic background fluid [23–29], but has recently been studied in the context of modified gravity theories as well [27,28,30]. Although the model mimics early- and late-time cosmic evolution scenarios in the asymptotic limits, certain technical issues remain unresolved for it to be the final candidate for a dark fluid, such as the issues of large-scale structure formation and the violation of causality due to a negative speed of sound associated with the fluid.

[a] e-mail: shambel.sahlu@wku.edu.et

In this work, our aim is to reconstruct $f(T)$ gravity models from the different variants of the CG model, namely the original CG (OCG), the Generalized CG (GCG) and the modified generalized CG (MGCG) models. The idea is that, in that way, we will be able to mimic the same expansion behavior (as the CG) with a theory of gravity, in this case $f(T)$, but without having to worry about the exoticity of the fluid with the above-mentioned problems. To do this, we assume the usual torsion fluid description with effective pressure $p_T$ and energy density $\rho_T$ related via the CG equation of state. After obtaining the reconstructed $f(T)$ gravity models, we study the expansion dynamics of the Universe by calculating the energy density, pressure and equation of state parameter for the torsion fluid in the non-interacting fluid systems such as vacuum-torsion, radiation-torsion, dust-torsion and radiation-dust-torsion systems for the three variants of the CG model.

The layout of the manuscript is as follows: in the following section, we review the cosmology of $f(T)$ gravity together with the simplest CG cosmological models and relate the two through some sort of a master equation, which we then solve in Sect. (2.1) for different cases. In Sects. (2.2) and (2.3), we go a step or two further and reconstruct different $f(T)$ gravity models from the GCG and MGCG models, respectively. In Sect. 3, we use an alternative approach by considering the simplest CG models to obtain Lagrangian densities $f(T)$ that depend explicitly as functions of the torsion scalar $T$ only. We then devote Sect. (4) for discussions of our results and the conclusions.

## 2 The Chaplygin gas as a model of modified teleparallel gravitation

The modified Einstein field equations of $f(T)$ gravity with a clear analogy to GR equations are given by [31–33]

$$f'G_{ab} + \frac{1}{2}g_{ab}[f - f'T] - f''S_{ab}{}^c\nabla_c T = \kappa^2 \mathscr{T}_{ab}, \quad (1)$$

where $f' \equiv df/dT$, $f'' \equiv d^2f/dT^2$, $\mathscr{T}_{ab}$ denotes the usual EMT of the matter fluid expressed as

$$\mathscr{T}_a{}^b = \frac{1}{e}\frac{\delta(eL_m)}{\delta e^b_a}, \quad (2)$$

and the coupling constant $\kappa^2 \equiv \frac{8\pi G}{c^4}$. The above field equations can be re-written in a more compact form as

$$G_{ab} = \mathscr{T}_{ab}^T + \mathscr{T}_{ab}^{(m)}, \quad (3)$$

where we have defined the EMT of the torsion ($T$) fluid as [31]

$$\mathscr{T}_{ab}^T = -\frac{1}{2f'}g_{ab}(f - f'T) - \frac{1}{f'}(f''S_{ab}{}^d\nabla_d T) \\ - \frac{1}{f'}(f' - 1)\mathscr{T}_{ab}^{(m)}. \quad (4)$$

In the limiting case of $f(T) = T$ (cf. [31,32]) the field equations reduce to those of GR.[1] From this generalized form of the gravitational field equations of motion in $f(T)$ gravity, we obtain the modified Friedmann and Raychaudhuri equations in $f(T)$ for FLRW spacetimes as follows[2]:

$$H^2 = \frac{\rho_m}{3f'} - \frac{1}{6f'}(f - Tf'), \quad (5)$$

$$2\dot{H} + 3H^2 = \frac{p_m}{f'} + \frac{1}{2f'}(f - Tf') + \frac{4f''H\dot{T}}{f'}, \quad (6)$$

where $H(t)$ is the Hubble (expansion) parameter defined from the scale factor $a(t)$ and the cosmic time $t$ as $H \equiv \frac{\dot{a}}{a}$. One can compute the torsion contributions of the thermodynamical quantities such as energy density $\rho_T$ and pressure $p_T$ from the EMT of the torsion fluid as follows:

$$\rho_T = -\frac{1}{f'}\left[(f' - 1)\rho_m + \frac{1}{2}(f - Tf')\right], \quad (7)$$

$$p_T = -\frac{1}{f'}\left[(f' - 1)p_m - \frac{1}{2}(f - Tf')\right] + 2H\frac{f''}{f'}\dot{T}, \quad (8)$$

where $\rho_m$ is the energy density, $p_m$ is the pressure of the matter fluid, torsion scalar $T = -6H^2$. Here we assume a slowly changing torsion fluid *i.e.*, $\dot{T} \approx 0$ such that the pressure of the torsion fluid is given by

$$p_T = -\frac{1}{f'}\left[(f' - 1)p_m - \frac{1}{2}(f - Tf')\right]. \quad (9)$$

Since the effective energy density of the fluid $\rho_{eff}$ is the sum of the two non-interacting fluid components (matter and torsion), we have the effective (total) energy density

$$\rho_{eff} \equiv \rho_m + \rho_T, \quad (10)$$

and the effective pressure of the total fluid is

$$p_{eff} \equiv p_m + p_T. \quad (11)$$

Then, the corresponding conservation equation of the fluid is

$$\dot{\rho} \equiv -3H(\rho_{eff} + p_{eff}). \quad (12)$$

---

[1] Here we assume geometric units where $\kappa = 1 = 8\pi G = c$, where $c$ is the speed of light and we use the $(+, -, -, -)$ metric convention for this manuscript.

[2] Overhead dots denote differentiation w.r.t. cosmic time.

From the modified Friedmann equation (5), the effective energy density of the total fluid is

$$\rho_{eff} = \frac{\rho_m}{f'} - \frac{1}{2f'}(f - Tf'), \quad (13)$$

and the corresponding effective pressure takes the form

$$p_{eff} = \frac{p_m}{f'} + \frac{1}{2f'}(f - Tf'). \quad (14)$$

In the special case $f(T) = T$, Eqs. (1)–(14) all reduce to the GR limit, which in turn describes similar cosmic dynamics as GR. A cosmological fluid of particular interest in recent years is the so-called CG. It is a fluid model proposed as a candidate for a unified description of dark matter and dark energy [4,29,34], and its cosmological scenarios are widely presented in the literature. In this work, we consider the torsion fluid as an exotic fluid with equations of state similar to the ones for three variants of the CG model and we reconstruct $f(T)$ gravity toy models corresponding to the original, generalized and modified generalized models. The characteristic equation of state for the CG model is given by [4,35]

$$p = -\frac{A}{\rho^\alpha}, \quad (15)$$

where $0 < \alpha \leq 1$, $A$ is a positive constant and the energy density $\rho > 0$.

Now, if we consider the possibility of the torsion fluid mimicking the CG with the characteristic equation of state given by

$$p_T = -\frac{A}{\rho_T^\alpha}, \quad (16)$$

and substitute the energy density $\rho_T$ and pressure $p_T$ of the torsion fluid from Eqs. (7) and (9) into Eq. (16), we obtain the master equation

$$-\frac{1}{f'}\left[(f'-1)p_m - \frac{1}{2}(f-Tf')\right]$$
$$\left[-\frac{1}{f'}\left((f'-1)\rho_m + \frac{1}{2}(f-Tf')\right)\right]^\alpha = -A, \quad (17)$$

from which our solution process starts. This is a general expression of the equation of state to reconstruct different $f(T)$ gravity models from the given two paradigmatic models of CG (original and generalized) and in the following two sections we reconstruct different $f(T)$ gravity models based on Eq. (17). As a consequence of field equation (1) and the EMT of the torsion fluid Eq. (4), the Lagrangian density of $f(T)$ gravity for different systems may depend on the torsion and matter fluids. A similar way of $f(T)$ representation is done in [36].

Let us discuss the scheme that we follow in the original and generalized CG models. We first reconstruct the $f(T)$ gravity model for five different cases: vacuum, radiation-torsion, dust-torsion, stiff matter-torsion and radiation-dust-torsion systems for both CG models. Then, secondly, we substitute the reconstructed $f(T)$ gravity models into Eqs. (7) and (9) and compute the corresponding energy density, pressure and the equation of state parameter of the torsion fluid for each case in these CG models. The energy density of each fluid apart from that of CG is expressed as

$$\rho_d = \rho_{d0} a^{-3},$$
$$\rho_r = \rho_{r0} a^{-4}, \text{ and}$$
$$\rho_s = \rho_{s0} a^{-6},$$

where $\rho_d$, $\rho_r$ and $\rho_s$ are the energy density for dust, radiation and stiff matter fluids respectively and $\rho_{i0}$ denotes the present-day value of the energy density of fluid type $i = \{d, r, s\}$.[3] Consequently, we compute the effective energy density and pressure of the effective fluid as well, the equation of state parameter for torsion fluid $w_T$ and we present the numerical plots of the effective equation state parameters $w_{eff(j)}$ for all total fluids as

$$w_{eff(j)} = \frac{p_T^j + p^i}{\rho_T^j + \rho^i}, \quad (18)$$

where the indice $j \equiv \{1, 2, 3, \ldots N\}$ depends on the number of reconstructed $f(T)$ gravity solutions, because we have more than one reconstructed $f(T)$ gravity models for each CG model. Thirdly, we define the parameter $\xi_j$ to represent the growth factor of the energy densities of the torsion fluid as

$$\xi_j = \frac{\rho_T^j}{\rho_{T0}^j}, \quad (19)$$

where $\rho_{T0}$ represents the energy density of the torsion in the present day. We also define another dimensionless parameter $\chi_j$ to represent the fraction of the effective energy densities of the total fluid from Eq. (10) as

$$\chi_j = \frac{\rho_{eff(j)}}{\rho_{eff0(j)}}, \quad (20)$$

where $\rho_{eff0}$ denotes the energy density of the total fluid in the present day for radiation-torsion, dust-torsion and radiation-dust-torsion systems in each CG model. Then, we present the numerical plots of ξ and χ versus the scale factor $a$ for radiation-torsion, dust-torsion and radiation-dust-torsion systems. This is the set of procedures we follow to reconstruct different $f(T)$ gravity models and the corresponding thermodynamic quantities accordingly. Similar procedures apply for the MGCG model, the only difference being the modified equation of state Eq. (168) we will use instead of Eq. (17). Finally, we have put a generalized discuss for all evolution of the equation of state parameters and the fractional energy densities in Sect. (4).

---

[3] For illustrative purpose we use:
1. $\rho_{r,0} = 3H_0^2 \Omega_{r,0}$, where $H_0 = 67.66$ km/s/Mpc [37] and $\Omega_{r,0} = 8.48 \times 10^{-5}$ [38],
2. $\rho_{d,0} = 3H_0^2 \Omega_{d,0}$ where $\Omega_{d,0} = 0.321$ [37] and
3. $\rho_{s,0} = 3H_0^2 \Omega_{s,0}$ where $\Omega_{s,0} = 10^{-6}$ [38].

### 2.1 Reconstructing modified teleparallel gravity from the OCG model

In the OCG model, the equation of state Eq. (15) has $\alpha = 1$ [24,39], and with the torsion fluid acting as the exotic CG fluid, we will have

$$p_T = -\frac{A}{\rho_T}. \tag{21}$$

In the following, different $f(T)$-gravity models will be reconstructed for vacuum, radiation-torsion, dust-torsion, stiff matter-torsion and radiation-dust-torsion systems in the OCG model.

#### 2.1.1 Vacuum system

In this case we assume the energy density and the pressure of the matter fluid are negligible, $\rho_m = p_m = 0$ and that torsion manifests itself as a CG. From the general expression of Eq. (17), we obtain

$$\left(T^2 - 4A\right) f'^2 - 2fTf' + f^2 = 0, \tag{22}$$

thus obtaining two possible $f(T)$ gravity models as solutions:

$$f_1(T) = c\left(T - 2\sqrt{A}\right), \tag{23}$$
$$f_2(T) = c\left(T + 2\sqrt{A}\right), \tag{24}$$

where $c \neq 0$ and it is an integration constant. By substituting $f_1(T)$ and $f_2(T)$ into Eq. (7), the energy density of the torsion fluid becomes

$$\rho_T = \pm\sqrt{A}. \tag{25}$$

As we indicated earlier in Eq. (15), the energy density of this exotic fluid is always positive and different from zero. Then, we take only the positive term of Eq. (25) $\rho_T = \sqrt{A}$, resulting in a negative pressure. This energy density agrees with the result in [34]. Consequently, we only take the negative term from the reconstructed $f(T)$ gravity, $f(T) = cT - 2c\sqrt{A}$, such that $p_T = -\sqrt{A}$, and the equation of state parameter of the torsion fluid $w_T = p_T/\rho_T = -1$. In the case of $f(T) = T$, the $f(T)$ gravity theory coincides with GR. Here, we observe that the obtained equation of state parameter of torsion asymptotically approaches the DE phase $w = -1$. This indicates that the torsion fluid acts as an exotic fluid and it is an alternative approach to describe the accelerated expansion of the Universe.

#### 2.1.2 Radiation-torsion system

Here we reconstruct torsion-radiation coupling by considering a non-interacting two-component fluid system of the Universe such that Eq. (17) is given as

$$-\frac{1}{f'}\left[(f'-1)\frac{1}{3}\rho_r - \frac{1}{2}(f - Tf')\right]$$
$$\left[-\frac{1}{f'}\left((f'-1)\rho_r + \frac{1}{2}(f - Tf')\right)\right] = -A. \tag{26}$$

From this equation we obtain four different $f(T)$ gravity models as follows:

$$f_1(T) = Tc - 2/3\rho_r c + 2/3\rho_r + 2/3\sqrt{4c^2\rho_r^2 + 9Ac^2 - 8c\rho_r^2 + 4\rho_r^2}, \tag{27}$$
$$f_2(T) = Tc - 2/3\rho_r c - 2/3, \rho_r + 2/3\sqrt{4c^2\rho_r^2 + 9Ac^2 - 8, c\rho_r^2 + 4\rho_r^2}, \tag{28}$$
$$f_3(T) = 2\frac{\left(2T\rho_r + 3A + \sqrt{-9AT^2 + 12AT\rho_r + 12A\rho_r^2 + 36A^2}\right)\rho_r}{4\rho_r^2 + 9A}, \tag{29}$$
$$f_4(T) = 2\frac{\left(2T\rho_r + 3A - \sqrt{-9AT^2 + 12AT\rho_r + 12A\rho_r^2 + 36A^2}\right)\rho_r}{4\rho_r^2 + 9A}. \tag{30}$$

In some limiting cases, where $\rho_r = 0$, $f_1(T)$ and $f_2(T)$ in Eqs. (27) and (28) are reduced to the vacuum system in Eqs. (23) and (24) respectively, while other $f(T)$ gravity models such as $f_3(T)$ and $f_4(T)$ in Eqs. (29) and (30) go to zero. By substituting Eqs. (27)–(30) into Eq. (7),we bring in the corresponding energy density of the torsion fluid as follows:

$$\rho_T^1 = \frac{-2\rho_r c + 2\rho_r + \sqrt{4(c-1)^2 \rho_r^2 + 9Ac^2}}{3c}, \tag{31}$$

$$\rho_T^2 = 1/3 \frac{-2\rho_r c + 2\rho_r - \sqrt{4(c-1)^2 \rho_r^2 + 9Ac^2}}{c}, \tag{32}$$

$$\rho_T^3 = -3 \frac{\left(\left(\frac{2}{3}\rho_r T - \frac{4}{3}\rho_r^2 - 2A\right)\sqrt{3} + \sqrt{A(-3T^2 + 4\rho_r T + 4\rho_r^2 + 12A)}\right)A}{3\sqrt{3}A(T - 2/3\rho_r) - 2\rho_r \sqrt{A(-3T^2 + 4\rho_r T + 4\rho_r^2 + 12A)}}, \tag{33}$$

$$\rho_T^4 = 3 \frac{A\left(\left(-2/3\rho_r T + 4/3\rho_r^2 + 2A\right)\sqrt{3} + \sqrt{A(-3T^2 + 4\rho_r T + 4\rho_r^2 + 12A)}\right)}{2\rho_r \sqrt{A(-3T^2 + 4\rho_r T + 4\rho_r^2 + 12A)} + 3\sqrt{3}A(T - 2/3\rho_r)}. \tag{34}$$

In a similar manner, we present the corresponding pressure of the torsion fluid in the era of radiation by substituting Eqs. (27)–(30) into Eq. (9). So, the reconstructed pressures of the torsion fluid are given as follows:

$$p_T^1 = \frac{-4\rho_r c + 4\rho_r - \sqrt{4(c-1)^2 \rho_r^2 + 9Ac^2}}{3c}, \tag{35}$$

$$p_T^2 = 1/3 \frac{-4\rho_r c + 4\rho_r + \sqrt{4(c-1)^2 \rho_r^2 + 9Ac^2}}{c}, \tag{36}$$

$$p_T^3 = -6 \frac{A\left(-\sqrt{A(-3T^2 + 4\rho_r T + 4\rho_r^2 + 12A)} + \sqrt{3}(2/3\rho_r T + A)\right)}{2\rho_r \sqrt{A(-3T^2 + 4\rho_r T + 4\rho_r^2 + 12A)} + 3\sqrt{3}A(T - 2/3\rho_r)}, \tag{37}$$

$$p_T^4 = -6 \frac{A\left(\sqrt{A(-3T^2 + 4\rho_r T + 4\rho_r^2 + 12A)} + \sqrt{3}(2/3\rho_r T + A)\right)}{-2\rho_r \sqrt{A(-3T^2 + 4\rho_r T + 4\rho_r^2 + 12A)} + 3\sqrt{3}A(T - 2/3\rho_r)}. \tag{38}$$

Therefore from Eqs. (31)–(34) and Eqs. (35)–(38) we also present the equation of state parameters of the torsion fluid accordingly:

$$w_T^1 = \frac{-4\rho_r c + 4\rho_r - \sqrt{4(c-1)^2 \rho_r^2 + 9Ac^2}}{-2\rho_r c + 2\rho_r + \sqrt{4(c-1)^2 \rho_r^2 + 9Ac^2}}, \tag{39}$$

$$w_T^2 = \frac{4\rho_r c - \sqrt{4(c-1)^2 \rho_r^2 + 9Ac^2} - 4\rho_r}{2\rho_r c + \sqrt{4(c-1)^2 \rho_r^2 + 9Ac^2} - 2\rho_r}, \tag{40}$$

$$w_T^3 = \frac{6\sqrt{A(-3T^2 + 4\rho_r T + 4\rho_r^2 + 12A)} + (-4\rho_r T - 6A)\sqrt{3}}{3\sqrt{A(-3T^2 + 4\rho_r T + 4\rho_r^2 + 12A)} + (-2\rho_r T + 4\rho_r^2 + 6A)\sqrt{3}}, \tag{41}$$

$$w_T^4 = \frac{-6\sqrt{A(-3T^2 + 4\rho_r T + 4\rho_r^2 + 12A)} + (-4\rho_r T - 6A)\sqrt{3}}{-3\sqrt{A(-3T^2 + 4\rho_r T + 4\rho_r^2 + 12A)} + (-2\rho_r T + 4\rho_r^2 + 6A)\sqrt{3}}. \tag{42}$$

We apply the definition of Eq. (18) and we represent the numerical plots of the equation of state parameter for the

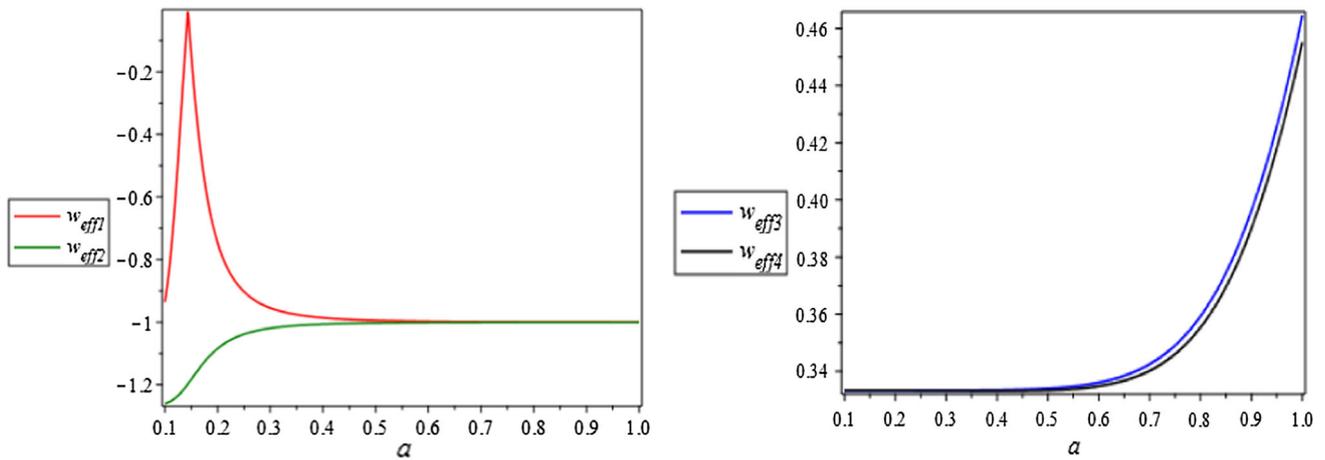

**Fig. 1** Both panels show the plots of $w_{eff(1,2,3,4)}$ versus $a$ in the case of a radiation-torsion system for OCG at $c = 2.5$ and $A = 1$

radiation-torsion system in the following figure for OCG Fig. 1.[4]

Then, the growth factor of the energy density of the torsion fluid in the radiation dominated era from the above thermodynamic quantities become:

$$\xi_1 = \frac{\rho_T^1}{\rho_{T_0}^1}, \quad \xi_2 = \frac{\rho_T^2}{\rho_{T_0}^2}, \quad \xi_3 = \frac{\rho_T^3}{\rho_{T_0}^3}, \quad \xi_4 = \frac{\rho_T^4}{\rho_{T_0}^4}. \tag{43}$$

And from Eqs. (31)–(34), we obtain the growth factor of the energy density for effective fluid as follows:

$$\xi_1 = \frac{-2\rho_r c + 2\rho_r + \sqrt{4(c-1)^2 \rho_r^2 + 9Ac^2}}{-2\rho_{r,0} c + 2\rho_{r,0} + \sqrt{4(c-1)^2 \rho_{r,0}^2 + 9Ac^2}}, \tag{44}$$

$$\xi_2 = \frac{-2\rho_r c + 2\rho_r - \sqrt{4(c-1)^2 \rho_r^2 + 9Ac^2}}{-2\rho_{r,0} c + 2\rho_{r,0} - \sqrt{4(c-1)^2 \rho_{r,0}^2 + 9Ac^2}}, \tag{45}$$

$$\xi_3 = \frac{-3\frac{\left((2/3\rho_r T - 4/3\rho_r^2 - 2A)\sqrt{3} + \sqrt{A(-3T^2 + 4\rho_r T + 4\rho_r^2 + 12A)}\right)A}{3\sqrt{3}A(T - 2/3\rho_r) - 2\rho_r\sqrt{A(-3T^2 + 4\rho_r T + 4\rho_r^2 + 12A)}}}{-3\frac{\left((2/3\rho_{r,0} T - 4/3\rho_{r,0}^2 - 2A)\sqrt{3} + \sqrt{A(-3T^2 + 4\rho_{r,0} T + 4\rho_{r,0}^2 + 12A)}\right)A}{3\sqrt{3}A(T - 2/3\rho_{r,0}) - 2\rho_{r,0}\sqrt{A(-3T^2 + 4\rho_{r,0} T + 4\rho_{r,0}^2 + 12A)}}}, \tag{46}$$

$$\xi_4 = \frac{-3\frac{\left((2/3\rho_r T - 4/3\rho_r^2 - 2A)\sqrt{3} - \sqrt{A(-3T^2 + 4\rho_r T + 4\rho_r^2 + 12A)}\right)A}{3\sqrt{3}A(T - 2/3\rho_r) + 2\rho_r\sqrt{A(-3T^2 + 4\rho_r T + 4\rho_r^2 + 12A)}}}{-3\frac{\left((2/3\rho_{r,0} T - 4/3\rho_{r,0}^2 - 2A)\sqrt{3} - \sqrt{A(-3T^2 + 4\rho_{r,0} T + 4\rho_{r,0}^2 + 12A)}\right)A}{3\sqrt{3}A(T - 2/3\rho_{r,0}) + 2\rho_{r,0}\sqrt{A(-3T^2 + 4\rho_{r,0} T + 4\rho_{r,0}^2 + 12A)}}}. \tag{47}$$

Here we also present the other dimensionless parameter $\chi$ to represent the fraction of the effective energy density of the radiation-torsion fluid:

$$\chi_1 = \frac{\rho_{eff}^1}{\rho_{eff_0}^1}, \quad \chi_2 = \frac{\rho_{eff}^2}{\rho_{eff_0}^2},$$

$$\chi_1 = \frac{\rho_{eff}^3}{\rho_{eff_0}^3}, \quad \chi_4 = \frac{\rho_{eff}^4}{\rho_{eff_0}^4}, \tag{48}$$

the explicit values of which are given by:

$$\chi_1 = \frac{3c\rho_r + -2\rho_r c + 2\rho_r + \sqrt{4(c-1)^2\rho_r^2 + 9Ac^2}}{3c\rho_{r,0} + -2\rho_{r,0} c + 2\rho_{r,0} + \sqrt{4(c-1)^2\rho_{r,0}^2 + 9Ac^2}}, \tag{49}$$

$$\chi_2 = \frac{3c\rho_r + -2\rho_r c + 2\rho_r - \sqrt{4(c-1)^2\rho_r^2 + 9Ac^2}}{3c\rho_{r,0} + -2\rho_{r,0} c + 2\rho_{r,0} - \sqrt{4(c-1)^2\rho_{r,0}^2 + 9Ac^2}}, \tag{50}$$

$$\chi_3 = \frac{\rho_r - 3\frac{\left((2/3\rho T - 4/3\rho_r^2 - 2A)\sqrt{3} + \sqrt{A(-3T^2 + 4\rho_r T + 4\rho_r^2 + 12A)}\right)A}{3\sqrt{3}A(T - 2/3\rho_r) - 2\rho_r\sqrt{A(-3T^2 + 4\rho_r T + 4\rho_r^2 + 12A)}}}{\rho_{r,0} - 3\frac{\left((2/3\rho_{r,0} T - 4/3\rho_{r,0}^2 - 2A)\sqrt{3} + \sqrt{A(-3T^2 + 4\rho_{r,0} T + 4\rho_{r,0}^2 + 12A)}\right)A}{3\sqrt{3}A(T - 2/3\rho_{r,0}) - 2\rho_{r,0}\sqrt{A(-3T^2 + 4\rho_{r,0} T + 4\rho_{r,0}^2 + 12A)}}}, \tag{51}$$

$$\chi_4 = \frac{\rho_r - 3\frac{\left((2/3\rho_r T - 4/3\rho_r^2 - 2A)\sqrt{3} - \sqrt{A(-3T^2 + 4\rho_r T + 4\rho_r^2 + 12A)}\right)A}{3\sqrt{3}A(T - 2/3\rho_r) + 2\rho_r\sqrt{A(-3T^2 + 4\rho_r T + 4\rho_r^2 + 12A)}}}{\rho_{r,0} - 3\frac{\left((2/3\rho_{r,0} T - 4/3\rho_{r,0}^2 - 2A)\sqrt{3} - \sqrt{A(-3T^2 + 4\rho_{r,0} T + 4\rho_{r,0}^2 + 12A)}\right)A}{3\sqrt{3}A(T - 2/3\rho_{r,0}) + 2\rho_{r,0}\sqrt{A(-3T^2 + 4\rho_{r,0} T + 4\rho_{r,0}^2 + 12A)}}}. \tag{52}$$

All the above thermodynamic quantities namely $\rho_T$, $p_T$, $w_T$ and $\rho_{eff}$ are depend on the energy density of the radiation fluid $\rho_r$ and $\rho_r$ is proportional to the cosmological scale factor. Consequently, the energy density parameters such as $\xi$ and $\chi$ also depend the scale factor of the Universe. One can obtain the growth factor of the energy densities of the fluids $\xi$ and $\chi$; here we present numerical results in Fig. 2.

### 2.1.3 Dust-torsion system

After the era of radiation, the Universe was predominantly filled by pressureless matter (dust) fluid. Thus, in a dust-dominated Universe, $p_d \approx 0$, and the equation of state

---
[4] In this paper, we consider that $0 < c$ in all models for each cases.

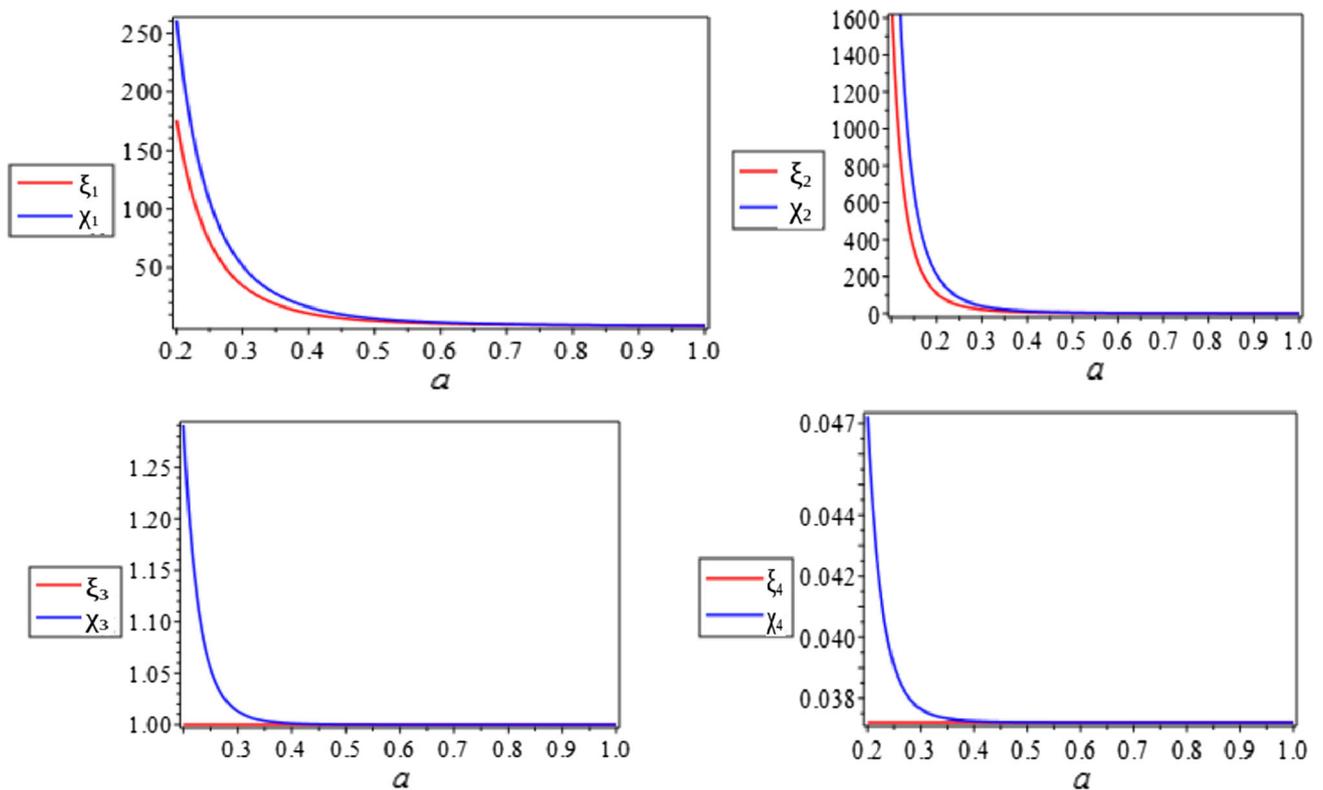

**Fig. 2** For the plots: in the upper-left panel, $\xi_1$ associated with $\chi_1$ versus $a$, in the upper-right panel, $\xi_2$, $\chi_2$ versus $a$, in the lower-left panel, $\xi_3$, $\chi_3$ versus $a$ and in the lower-right panel, $\xi_4$, $\chi_4$ versus $a$ for radiation- torsion system in OCG. We use $c = 0.9$ and $A = 0.7$ for the numerical plotting

parameter is $w_d \approx 0$. In our current model, we consider the non-interacting dust-torsion system and briefly discuss its implications for cosmic expansion. The general expression of Eq. (17) in the dust-torsion system becomes

$$\frac{1}{f'^2}\left[\frac{1}{4}(f - Tf')^2 + \frac{1}{2}(f' - 1)(f - Tf')\rho_d\right] = A, \quad (53)$$

and this equation admits four different $f(T)$ gravity models as solutions:

$$f_1(T) = (1 - c)\rho_d + cT + \sqrt{(c^2 - 2c)\rho_d^2 + 4c_1^2 A}, \quad (54)$$

$$f_2(T) = (1 - c)\rho_d + cT - \sqrt{(c^2 - 2c)\rho_d^2 + 4c_1^2 A}, \quad (55)$$

$$f_3(T) = \frac{\rho_d}{\rho_d^2 + 4A}\left[\rho_d T + 4A - 2\sqrt{-AT^2 + 2AT\rho_d + 4A^2}\right], \quad (56)$$

$$f_4(T) = \frac{\rho_d}{\rho_d^2 + 4A}\left[\rho_d T + 4A + 2\sqrt{-AT^2 + 2AT\rho_d + 4A^2}\right]. \quad (57)$$

For $\rho_d = 0$, two of the solutions, Eqs. (54) and (55), reduce to the vacuum case, Eqs. (23) and (24) respectively, whereas the other two solutions, Eqs. (56) and (57), vanish. Let us now substitute $f_1(T)$, $f_2(T)$, $f_3(T)$ and $f_4(T)$ into Eqs. (7) and (9) to obtain the energy densities of the torsion

$$\rho_T^1 = \frac{-\rho_d c + \rho_d + \sqrt{(c-1)^2 \rho_d^2 + 4Ac^2}}{2c}, \quad (58)$$

$$\rho_2^T = \frac{-\rho_d c + \rho_d - \sqrt{(c-1)^2 \rho_d^2 + 4Ac^2}}{2c}, \quad (59)$$

$$\rho_T^3 = \frac{A\left(-\rho_d T + 2\rho_d^2 + 4A + 2\sqrt{4A^2 - T(T - 2\rho_d)A}\right)}{\rho_d\sqrt{4A^2 - T(T - 2\rho_d)A} + 2A(T - \rho_d)}, \quad (60)$$

$$\rho_T^4 = \frac{A\left(-T\rho_d + 2\rho_d^2 + 2A - \sqrt{A(-T^2 + 4\rho_d^2 + 4A)}\right)}{AT + \rho_d\sqrt{A(-T^2 + 4\rho_d^2 + 4A)}}, \quad (61)$$

and the corresponding isotropic pressures

$$p_T^1 = \frac{-3\rho_d c + 3\rho_d - \sqrt{(c-1)^2 \rho_d^2 + 4Ac^2}}{2c}, \quad (62)$$

$$p_T^2 = \frac{-3c\rho_d + 3\rho_d + \sqrt{(c-1)^2 \rho_d^2 + 4Ac^2}}{2c}, \quad (63)$$

$$p_T^3 = \frac{A\left(-3\rho_d T + 2\rho_d^2 + 6\sqrt{4A^2 - T(T - 2\rho_d)A}4A\right)}{(2T - 2\rho_d)A + \rho_d\sqrt{4A^2 - T(T - 2\rho_d)A}}, \quad (64)$$

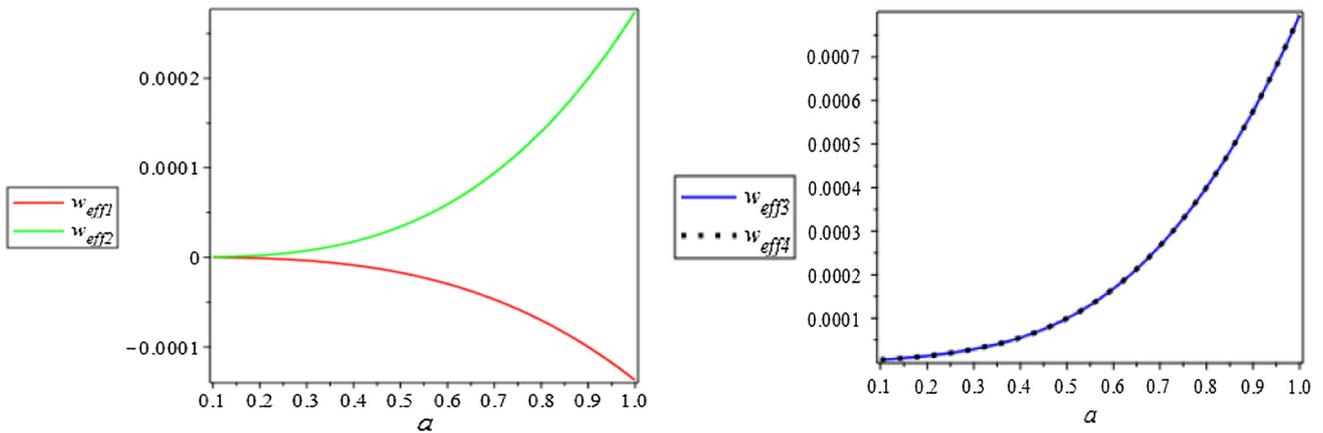

**Fig. 3** All plots show $w_{eff(1,2,3,4)}$ versus $a$ for dust-torsion system for OCG for $c = 2$ and $A = 0.98$

$$p_T^4 = -\frac{A\left(T\rho_d + 2\rho_d^2 + 2A - \sqrt{A\left(-T^2 + 4\rho_d^2 + 4A\right)}\right)}{AT + \rho_d\sqrt{A\left(-T^2 + 4\rho_d^2 + 4A\right)}}. \tag{65}$$

The reconstructed EoS parameters of the torsion fluid during the dust-dominated phase $w_T = p_T/\rho_T$ are then given by:

$$w_T^1 = \frac{-3\rho_d c + 3\rho_d - \sqrt{(c-1)^2\rho_d^2 + 4Ac^2}}{-\rho_d c + \rho_d + \sqrt{(c-1)^2\rho_d^2 + 4Ac^2}}, \tag{66}$$

$$w_T^2 = \frac{3c\rho_d - \sqrt{(c-1)^2\rho_d^2 + 4Ac^2} - 3\rho_d}{c\rho_d + \sqrt{(c-1)^2\rho_d^2 + 4Ac^2} - \rho_d}, \tag{67}$$

$$w_T^3 = \frac{-3\rho_d T + 2\rho_d^2 + 6\sqrt{4A^2 - T(T-2\rho_d)A}4A}{-\rho_d T + 2\rho_d^2 + 4A + 2\sqrt{4A^2 - T(T-2\rho_d)A}}, \tag{68}$$

$$w_T^4 = \frac{-T\rho_d - 2\rho_d^2 - 2A + \sqrt{A\left(-T^2 + 4\rho_d^2 + 4A\right)}}{-T\rho_d + 2\rho_d^2 + 2A + \sqrt{A\left(-T^2 + 4\rho_d^2 + 4A\right)}}. \tag{69}$$

We use Eq. (18) and represent the numerical plots of the equation of state parameter for the dust-torsion system in Fig. 3 figure for OCG.

Moreover, the fractional energy densities of torsion in the dust-torsion system give as follows:

$$\xi_1 = \frac{-\rho_d c + \rho_d + \sqrt{(c-1)^2\rho_d^2 + 4Ac^2}}{-\rho_{d,0} c + \rho_{d,0} + \sqrt{(c-1)^2\rho_{d,0}^2 + 4Ac^2}}, \tag{70}$$

$$\xi_2 = \frac{-\rho_d c + \rho_d - \sqrt{(c-1)^2\rho_d^2 + 4Ac^2}}{-\rho_{d,0} c + \rho_{d,0} - \sqrt{(c-1)^2\rho_{d,0}^2 + 4Ac^2}}, \tag{71}$$

$$\xi_3 = \frac{\frac{A\left(-\rho T + 2\rho^2 + 4A + 2\sqrt{4A^2 - T(T-2\rho)A}\right)}{\rho\sqrt{4A^2 - T(T-2\rho)A} + 2A(T-\rho)}}{\frac{A\left(-\rho_{d,0} T + 2\rho_{d,0}^2 + 4A + 2\sqrt{4A^2 - T(T-2\rho_{d,0})A}\right)}{\rho_{d,0}\sqrt{4A^2 - T(T-2\rho_{d,0})A} + 2A(T-\rho_{d,0})}}, \tag{72}$$

$$\xi_4 = \frac{\frac{A\left(-T\rho + 2\rho^2 + 2A - \sqrt{A(-T^2+4\rho^2+4A)}\right)}{AT + \rho\sqrt{A(-T^2+4\rho^2+4A)}}}{\frac{A\left(-T\rho_{d,0} + 2\rho_{d,0}^2 + 2A - \sqrt{A(-T^2+4\rho_{d,0}^2+4A)}\right)}{AT + \rho_{d,0}\sqrt{A(-T^2+4\rho_{d,0}^2+4A)}}}, \tag{73}$$

and for the effective fluid, these become:

$$\chi_1 = \frac{\rho_d c + \rho_d + \sqrt{(c-1)^2\rho_d^2 + 4Ac^2}}{\rho_{d,0} c + \rho_{d,0} + \sqrt{(c-1)^2\rho_{d,0}^2 + 4Ac^2}}, \tag{74}$$

$$\chi_2 = \frac{\rho_d c + \rho_d - \sqrt{(c-1)^2\rho_d^2 + 4Ac^2}}{\rho_{d,0} c + \rho_{d,0} - \sqrt{(c-1)^2\rho_{d,0}^2 + 4Ac^2}}, \tag{75}$$

$$\chi_3 = \frac{\rho_d + \frac{A\left(-\rho T + 2\rho^2 + 4A + 2\sqrt{4A^2 - T(T-2\rho)A}\right)}{\rho\sqrt{4A^2 - T(T-2\rho)A} + 2A(T-\rho)}}{\rho_{d,0} + \frac{A\left(-\rho_{d,0} T + 2\rho_{d,0}^2 + 4A + 2\sqrt{4A^2 - T(T-2\rho_{d,0})A}\right)}{\rho_{d,0}\sqrt{4A^2 - T(T-2\rho_{d,0})A} + 2A(T-\rho_{d,0})}}, \tag{76}$$

$$\chi_4 = \frac{\rho_d + \frac{A\left(-T\rho + 2\rho^2 + 2A - \sqrt{A(-T^2+4\rho^2+4A)}\right)}{AT + \rho\sqrt{A(-T^2+4\rho^2+4A)}}}{\rho_{d,0} + \frac{A\left(-T\rho_{d,0} + 2\rho_{d,0}^2 + 2A - \sqrt{A(-T^2+4\rho_{d,0}^2+4A)}\right)}{AT + \rho_{d,0}\sqrt{A(-T^2+4\rho_{d,0}^2+4A)}}}. \tag{77}$$

In Fig. 4, we present the growth of the fractional energy densities for torsion and effective fluids versus the scale factor.

### 2.1.4 Stiff matter-torsion system

Here, we consider a Universe composed of stiff matter ($p_s = \rho_s$ with $w = 1$) and torsion. The general expression of Eq. (17) for such a system is given as

$$-\frac{1}{f'}\left[(f'-1)\rho_s - \frac{1}{2}(f - Tf')\right]$$
$$\left[-\frac{1}{f'}\left((f'-1)\rho_s + \frac{1}{2}(f - Tf')\right)\right] = -A. \tag{78}$$

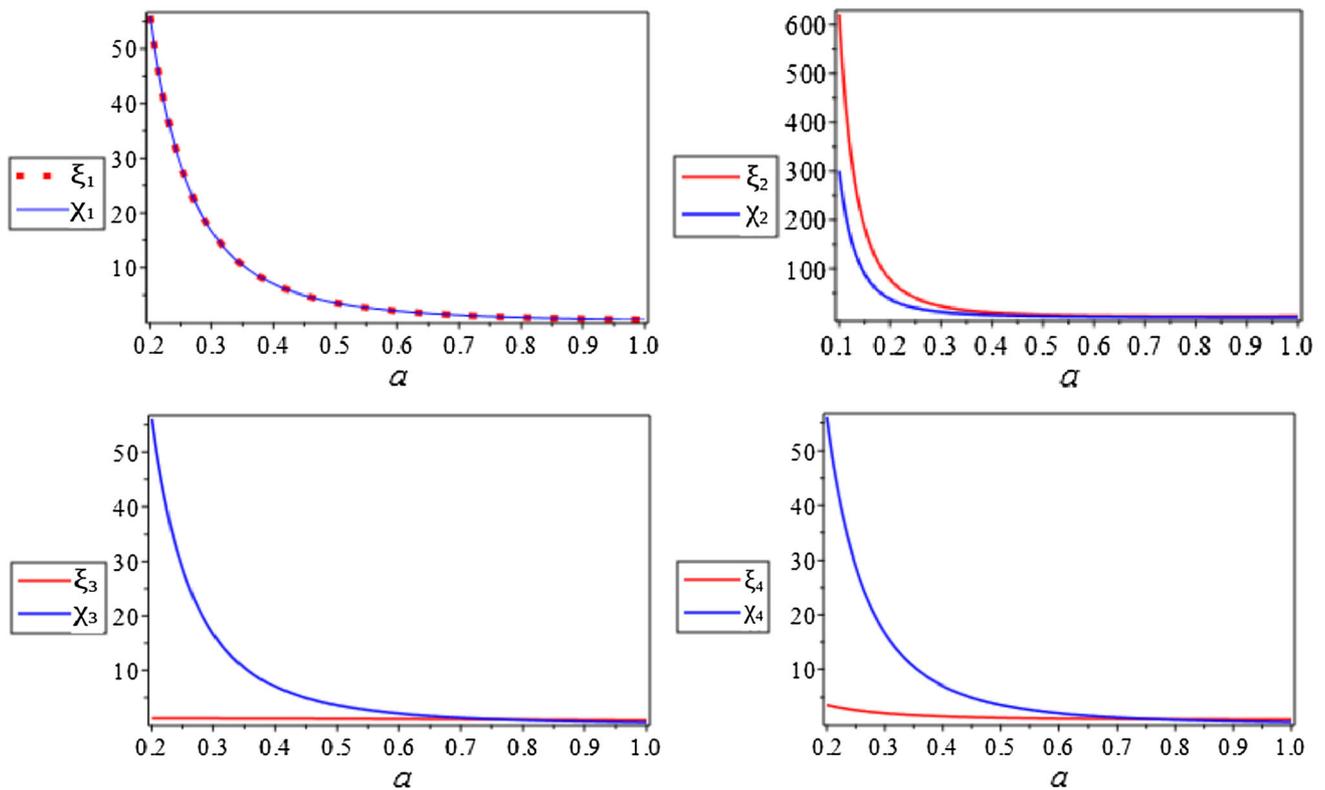

**Fig. 4** In all panels, we present $\xi_{1,2,3,4}$ associated with $\chi_{1,2,3,4}$ versus $a$ for dust-torsion system for OCG. We use $c = 1.9$ and $A = 0.5$ for these numerical plots

From this expression we reconstruct four different $f(T)$ gravity models as follows:

$$f_1(T) = Tc - 2\sqrt{c^2 \rho_s^2 + Ac^2 - 2c\rho_s^2 + \rho_s^2}, \quad (79)$$

$$f_2(T) = Tc + 2\sqrt{c^2 \rho_s^2 + Ac^2 - 2c\rho_s^2 + \rho_s^2}, \quad (80)$$

$$f_3(T) = \frac{\left(T\rho_s + \sqrt{-AT^2 + 4A\rho_s^2 + 4A^2}\right)\rho_s}{\rho_s^2 + A}, \quad (81)$$

$$f_4(T) = -\frac{\left(-T\rho_s + \sqrt{-AT^2 + 4A\rho_s^2 + 4A^2}\right)\rho_s}{\rho_s^2 + A}. \quad (82)$$

Here also we see that setting $\rho_s = 0$ reduces two of the $f(T)$ gravity models, Eqs. (79) and (80), to the vacuum cases, Eqs. (23) and (24) we studied earlier, whereas the other two solutions, Eqs. (81) and (82), both vanish. By substituting the above $f(T)$ solutions into Eq. (7), we obtain the corresponding energy densities and isotropic pressures of the torsion fluid in the era of stiff matter as follows:

$$\rho_T^1 = \frac{-\rho_s c + \rho_s + \sqrt{(c-1)^2 \rho_s^2 + Ac^2}}{c}, \quad (83)$$

$$\rho_T^2 = \frac{-\rho_s c + \rho_s - \sqrt{(c-1)^2 \rho_s^2 + Ac^2}}{c}, \quad (84)$$

$$\rho_T^3 = \frac{A\left(-T\rho_s + 2\rho_s^2 + 2A - \sqrt{A(-T^2 + 4\rho_s^2 + 4A)}\right)}{AT - \rho_s\sqrt{A(-T^2 + 4\rho_s^2 + 4A)}}, \quad (85)$$

$$\rho_T^4 = \frac{A\left(-T\rho_s + 2\rho_s^2 + 2A + \sqrt{A(-T^2 + 4\rho_s^2 + 4A)}\right)}{AT + \rho_s\sqrt{A(-T^2 + 4\rho_s^2 + 4A)}}, \quad (86)$$

$$p_T^1 = \frac{-\rho_s c + \rho_s - \sqrt{(c-1)^2 \rho_s^2 + Ac^2}}{c}, \quad (87)$$

$$p_T^2 = \frac{-c\rho_s + \rho_s + \sqrt{(c-1)^2 \rho_s^2 + Ac^2}}{c}, \quad (88)$$

$$p_T^3 = -\frac{A\left(T\rho_s + 2\rho_s^2 + 2A + \sqrt{A(-T^2 + 4\rho_s^2 + 4A)}\right)}{AT - \rho_s\sqrt{A(-T^2 + 4\rho_s^2 + 4A)}}, \quad (89)$$

$$p_T^4 = -\frac{A\left(T\rho_s + 2\rho_s^2 + 2A - \sqrt{A(-T^2 + 4\rho_s^2 + 4A)}\right)}{AT + \rho_s\sqrt{A(-T^2 + 4\rho_s^2 + 4A)}}. \quad (90)$$

We also compute the equation of state parameter of the torsion fluid in the stiff fluid-torsion system as follows:

$$w_T^1 = \frac{-\rho_s c + \rho_s - \sqrt{(c-1)^2 \rho_s^2 + Ac^2}}{-\rho_s c + \rho_s + \sqrt{(c-1)^2 \rho_s^2 + Ac^2}}, \tag{91}$$

$$w_T^2 = \frac{c\rho_s - \sqrt{(c-1)^2 \rho_s^2 + Ac^2} - \rho_s}{c\rho_s + \sqrt{(c-1)^2 \rho_s^2 + Ac^2} - \rho_s}, \tag{92}$$

$$w_T^3 = \frac{-T\rho_s - 2\rho_s^2 - 2A - \sqrt{A(-T^2 + 4\rho_s^2 + 4A)}}{-T\rho_s + 2\rho_s^2 + 2A - \sqrt{A(-T^2 + 4\rho_s^2 + 4A)}}, \tag{93}$$

$$w_T^4 = \frac{-T\rho_s - 2\rho_s^2 - 2A + \sqrt{A(-T^2 + 4\rho_s^2 + 4A)}}{-T\rho_s + 2\rho_s^2 + 2A + \sqrt{A(-T^2 + 4\rho_s^2 + 4A)}}. \tag{94}$$

### 2.1.5 Radiation-dust-torsion system

Here, we consider a non-interacting multi-fluid system, namely radiation and dust with torsion, and those fluids acts as an exotic fluid for cosmic expansion. In this context, all thermodynamical quantities are the mixture of the individual species. For instance, the pressure of the effective fluid is given as $p_{eff} = p_m + p_T$. The general form of Eq. (17) for radiation-dust-torsion system is given as

$$-\frac{1}{f'}\left[(f'-1)p - \frac{1}{2}(f - Tf')\right]$$
$$\left[-\frac{1}{f'}\left((f'-1)\rho + \frac{1}{2}(f - Tf')\right)\right] = -A, \tag{95}$$

where the energy density of matter $\rho = \rho_r + \rho_d$. We reconstruct four different $f(T)$ gravity models through this system as follows:

$$f_1(T) = -c\rho_d + Tc - 2/3\,c\rho_r + \rho_d + 2/3\,\rho_r$$
$$-1/3\Big[9\rho_d^2 c^2 + 24\,c^2\rho_d\rho_r + 16\,c^2\rho_r^2$$
$$+36\,Ac^2 - 18\,\rho_d^2 c - 48\,c\rho_d\rho_r - 32\,c\rho_r^2$$
$$+9\,\rho_d^2 + 24\,\rho_d\rho_r + 16\,\rho_r^2\Big], \tag{96}$$

$$f_2(T) = -c\rho_d + Tc - 2/3\,c\rho_r + \rho_d + 2/3\,\rho_r$$
$$+1/3\Big[9\,\rho_d^2 c^2 + 24\,c^2\rho_d\rho_r + 16\,c^2\rho_r^2$$
$$+36\,Ac^2 - 18\,\rho_d^2 c - 48\,c\rho_d\rho_r - 32\,c\rho_r^2$$
$$+9\,\rho_d^2 + 24\,\rho_d\rho_r + 16\,\rho_r^2\Big], \tag{97}$$

$$f_3(T) = \frac{1}{9\rho_d^2 + 24\rho_d\rho_r + 16\rho_r^2 + 36A}\Big\{9\rho_d^2 T$$
$$+24\rho_d\rho_r T + 16\rho_r^2 T + 36A\rho_d + 24A\rho_r$$
$$-2\Big[162A\rho_d^3 T + 108A\rho_d^3\rho_r - 81A\rho_d^2 T^2$$
$$+540A\rho_d^2 T\rho_r + 396A\rho_d^2\rho_r^2 - 216A\rho_d T^2\rho_r$$
$$+576A\rho_d T\rho_r^2 + 480A\rho_d\rho_r^3 - 144AT^2\rho_r^2$$
$$+192AT\rho_r^3 + 192A\rho_r^4 + 324A^2\rho_d^2$$
$$+864A^2\rho_d\rho_r + 576A^2\rho_r^2\Big]^{\frac{1}{2}}\Big\}, \tag{98}$$

$$f_4(T) = \frac{1}{9\rho_d^2 + 24\rho_d\rho_r + 16\rho_r^2 + 36A}\Big\{9\rho_d^2 T$$
$$+24\rho_d\rho_r T + 16\rho_r^2 T + 36A\rho_d + 24A\rho_r$$
$$+2\Big[162A\rho_d^3 T + 108A\rho_d^3\rho_r - 81A\rho_d^2 T^2$$
$$+540A\rho_d^2 T\rho_r + 396A\rho_d^2\rho_r^2 - 216A\rho_d T^2\rho_r$$
$$+576A\rho_d T\rho_r^2 + 480A\rho_d\rho_r^3 - 144AT^2\rho_r^2$$
$$+192AT\rho_r^3 + 192A\rho_r^4 + 324A^2\rho_d^2$$
$$+864A^2\rho_d\rho_r + 576A^2\rho_r^2\Big]^{\frac{1}{2}}\Big\}. \tag{99}$$

If $\rho_d = \rho_r = 0$, the above to $f(T)$ gravity models Eqs. (96) and (97) reduce to the vacuum case Eq. (23) and (24) respectively. The other two solutions Eqs. (98) and (99) go to zero in the vacuum limiting case. We substitute $f_1(T)$, $f_2(T)$, $f_3(T)$ and $f_4(T)$ into Eq. (7). Then, the energy density of the torsion fluid in radiation-dust-torsion system:

$$\rho_T^1 = \frac{-\varpi + (-3\rho_d - 4\rho_r)c + 3\rho_d + 4\rho_r}{6c}, \tag{100}$$

$$\rho_T^2 = \frac{\varpi + (-3\rho_d - 4\rho_r)c + 3\rho_d + 4\rho_r}{6c}, \tag{101}$$

$$\rho_T^3 = 2\frac{A\left(6(\rho_d + 4/3\rho_r)(2/3\rho_r^2 + (-T/3 + 7/6\rho_d)\rho_r - 1/4\rho_d T + 1/2\rho_d^2 + A)\sqrt{3} + \mathcal{N}\right)}{(\rho_d + 4/3\rho_r)\left(6A(T - \rho_d - 2/3\rho_r)\sqrt{3} + \mathcal{N}\right)}, \tag{102}$$

$$\rho_T^4 = 2\frac{A\left(-6(\rho_d + 4/3\rho_r)(2/3\rho_r^2 + (-T/3 + 7/6\rho_d)\rho_r - 1/4\rho_d T + 1/2\rho_d^2 + A)\sqrt{3} + \mathcal{N}\right)}{(\rho_d + 4/3\rho_r)\left(-6A(T - \rho_d - 2/3\rho_r)\sqrt{3} + \mathcal{N}\right)}, \tag{103}$$

where

$$\varpi = \sqrt{9(c-1)^2\rho_d^2 + 24\rho_r(c-1)^2\rho_d + 16(c-1)^2\rho_r^2 + 36Ac^2}$$

$$\mathcal{N} = \sqrt{(3\rho_d + 4\rho_r)^2 A(6\rho_d T + 4\rho_d\rho_r - 3T^2 + 4\rho_r T + 4\rho_r^2 + 12A)}.$$

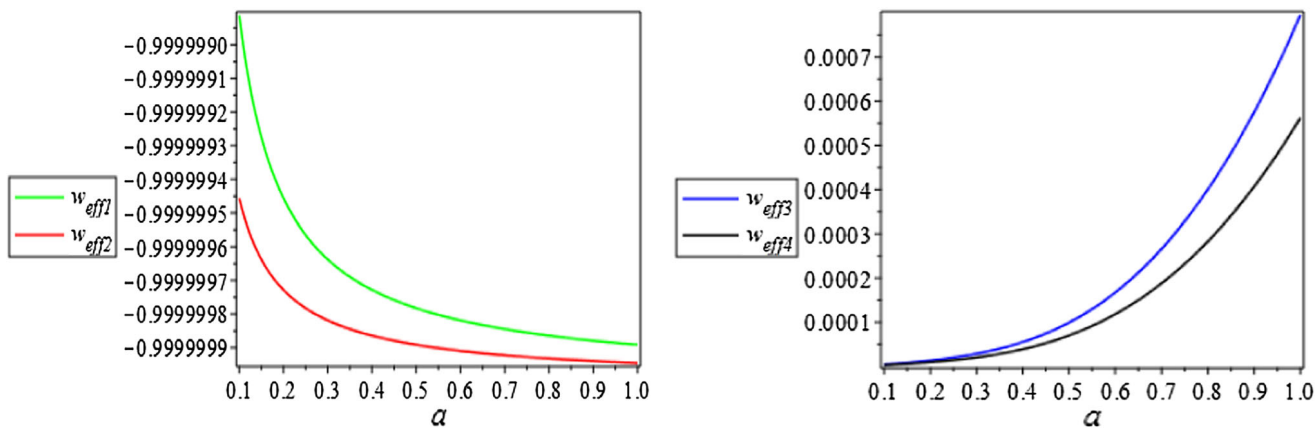

**Fig. 5** Both plots in the left and right panel show $w_{eff(1,2,3,4)}$ versus $a$ for radiation-dust-torsion system for OCG. We use $c = 4$ and $A = 1$ for plotting

And we also substitute $f_1(T)$, $f_2(T)$, $f_3(T)$ and $f_4(T)$ into Eq. (9), and reconstruct the correspond pressure of the fluid given as follows:

$$\rho_T = \frac{\varpi + (-9\rho_d - 8\rho_r)c + 9\rho_d + 8\rho_r}{6c}, \quad (104)$$

$$\rho_T = -\frac{\varpi + (-9\rho_d - 8\rho_r)c + 9\rho_d + 8\rho_r}{6c}, \quad (105)$$

$$p_T^3 = \frac{A - \left(2\left(\rho_d + \frac{4}{3}\rho_r\right)^2 \left(\left(\frac{2}{3}T - \rho_d/2\right)\rho_r + \frac{3}{4}\rho_d T - 1/2\rho_d^2 + A\right)\sqrt{3} + 1/9\mathcal{N}\left(9\rho_d + 8\rho_r\right)\right)}{6\left(\rho_d + 4/3\rho_r\right)^2 \left(-6A\left(T - \rho_d - \frac{2}{3}\rho_r\right)\sqrt{3} + \mathcal{N}\right)}, \quad (106)$$

$$p_T^4 = \frac{A\left(2\left(\rho_d + \frac{4}{3}\rho_r\right)^2 \left(\left(\frac{2}{3}T - \rho_d/2\right)\rho_r + \frac{3}{4}\rho_d T - 1/2\rho_d^2 + A\right)\sqrt{3} + 1/9\mathcal{N}\left(9\rho_d + 8\rho_r\right)\right)}{6\left(\rho_d + 4/3\rho_r\right)^2 \left(-6A\left(T - \rho_d - \frac{2}{3}\rho_r\right)\sqrt{3} + \mathcal{N}\right)}. \quad (107)$$

And the corresponding equation of state parameter of the torsion fluid express as follows:

$$w_T^1 = \frac{(9\rho_d + 8\rho_r)c + \varpi - 9\rho_d - 8\rho_r}{(3\rho_d + 4\rho_r)c - \varpi - 3\rho_d - 4\rho_r},$$

$$w_T^2 = \frac{(9\rho_d + 8\rho_r)c - \varpi - 9\rho_d - 8\rho_r}{(3\rho_d + 4\rho_r)c + \varpi - 3\rho_d - 4\rho_r},$$

$$w_T^3 = \frac{-108\left(\rho_d + 4/3\rho_r\right)^2 \left((2/3\,T - \rho_d/2)\rho_r + 3/4\,\rho_d T - 1/2\,\rho_d^2 + A\right)\sqrt{3} + 6\mathcal{N}}{18\left(\rho_d + 4/3\rho_r\right)\left(6\left(\rho_d + 4/3\rho_r\right)\left(2/3\,\rho_r^2 + (-T/3 + 7/6\,\rho_d)\rho_r - 1/4\,\rho_d T + 1/2\,\rho_d^2 + A\right)\sqrt{3} + \mathcal{N}\right)},$$

$$w_T^4 = \frac{108\left(\rho_d + 4/3\rho_r\right)^2 \left((2/3\,T - \rho_d/2)\rho_r + 3/4\,\rho_d T - 1/2\,\rho_d^2 + A\right)\sqrt{3} + 6\mathcal{N}}{18\left(\rho_d + 4/3\rho_r\right)\left(6\left(\rho_d + 4/3\rho_r\right)\left(2/3\,\rho_r^2 + (-T/3 + 7/6\,\rho_d)\rho_r - 1/4\,\rho_d T + 1/2\,\rho_d^2 + A\right)\sqrt{3} + \mathcal{N}\right)}.$$

We present the numerical plots of the evolution of effective equation of state parameter in Fig. 5 for non-interacting fluids (radiation-dust-torsion systems). Here we reconstruct the fraction of energy densities for torsion fluid in radiation-dust-torsion system as follows:

$$\xi_1 = \frac{-\varpi + (-3\rho_d - 4\rho_r)c + 3\rho_d + 4\rho_r}{-\varpi + (-3\rho_{d,0} - 4\rho_{r,0})c + 3\rho_{d,0} + 4\rho_{r,0}}, \quad (108)$$

$$\xi_2 = \frac{\sqrt{9(c-1)^2 \rho_d^2 + 24 \rho_r (c-1)^2 \rho_d + 16 (c-1)^2 \rho_r^2 + 36 A c^2} + (-3\rho_d - 4\rho_r)c + 3\rho_d + 4\rho_r}{\sqrt{9(c-1)^2 \rho_{d,0}^2 + 24 \rho_{r,0} (c-1)^2 \rho_{d,0} + 16 (c-1)^2 \rho_{r,0}^2 + 36 A c^2} + (-3\rho_{d,0} - 4\rho_{r,0})c + 3\rho_{d,0} + 4\rho_{r,0}},$$ (109)

$$\xi_3 = \frac{2\frac{A\left(6(\rho_d+4/3\rho_r)(2/3\rho_r^2+(-T/3+7/6\rho_d)\rho_r-1/4\rho_d T+1/2\rho_d^2+A)\sqrt{3}+\mathscr{N}\right)}{(\rho_d+4/3\rho_r)\left(6A(T-\rho_d-2/3\rho_r)\sqrt{3}+\mathscr{N}\right)}}{2\frac{A\left(6(\rho_{d,0}+4/3\rho_{r,0})(2/3\rho_{r,0}^2+(-T/3+7/6\rho_{d,0})\rho_{r,0}-1/4\rho_{d,0} T+1/2\rho_{d,0}^2+A)\sqrt{3}+\mathscr{N}\right)}{(\rho_{d,0}+4/3\rho_{r,0})\left(6A(T-\rho_{d,0}-2/3\rho_{r,0})\sqrt{3}+\mathscr{N}\right)}},$$ (110)

$$\xi_4 = \frac{2\frac{A\left(-6(\rho_d+4/3\rho_r)(2/3\rho_r^2+(-T/3+7/6\rho_d)\rho_r-1/4\rho_d T+1/2\rho_d^2+A)\sqrt{3}+\mathscr{N}\right)}{(\rho_d+4/3\rho_r)\left(-6A(T-\rho_d-2/3\rho_r)\sqrt{3}+\mathscr{N}\right)}}{2\frac{A\left(-6(\rho_{d,0}+4/3\rho_{r,0})(2/3\rho_{r,0}^2+(-T/3+7/6\rho_{d,0})\rho_{r,0}-1/4\rho_{d,0} T+1/2\rho_{d,0}^2+A)\sqrt{3}+\mathscr{N}\right)}{(\rho_{d,0}+4/3\rho_{r,0})\left(-6A(T-\rho_{d,0}-2/3\rho_{r,0})\sqrt{3}+\mathscr{N}\right)}},$$ (111)

$$\chi_1 = \frac{\mathscr{J} - \varpi + (-3\rho_d - 4\rho_r)c + 3\rho_d + 4\rho_r}{\mathscr{J} - \varpi + (-3\rho_{d,0} - 4\rho_{r,0})c + 3\rho_{d,0} + 4\rho_{r,0}},$$ (112)

$$\chi_2 = \frac{\mathscr{J} + \varpi + (-3\rho_d - 4\rho_r)c + 3\rho_d + 4\rho_r}{\mathscr{J} + \varpi + (-3\rho_{d,0} - 4\rho_{r,0})c + 3\rho_{d,0} + 4\rho_{r,0}},$$ (113)

$$\chi_3 = \frac{\rho_d + \rho_r + 2\frac{A\left(6(\rho_d+4/3\rho_r)(2/3\rho_r^2+(-T/3+7/6\rho_d)\rho_r-1/4\rho_d T+1/2\rho_d^2+A)\sqrt{3}+\mathscr{N}\right)}{(\rho_d+4/3\rho_r)\left(6A(T-\rho_d-2/3\rho_r)\sqrt{3}+\mathscr{N}\right)}}{\rho_{d,0} + \rho_{r,0} + 2\frac{A\left(6(\rho_{d,0}+4/3\rho_{r,0})(2/3\rho_{r,0}^2+(-T/3+7/6\rho_{d,0})\rho_{r,0}-1/4\rho_{d,0} T+1/2\rho_{d,0}^2+A)\sqrt{3}+\mathscr{N}\right)}{(\rho_{d,0}+4/3\rho_{r,0})\left(6A(T-\rho_{d,0}-2/3\rho_{r,0})\sqrt{3}+\mathscr{N}\right)}},$$ (114)

$$\chi_4 = \frac{\rho_d + \rho_r + 2\frac{A\left(-6(\rho_d+4/3\rho_r)(2/3\rho_r^2+(-T/3+7/6\rho_d)\rho_r-1/4\rho_d T+1/2\rho_d^2+A)\sqrt{3}+\mathscr{N}\right)}{(\rho_d+4/3\rho_r)\left(-6A(T-\rho_d-2/3\rho_r)\sqrt{3}+\mathscr{N}\right)}}{\rho_{d,0} + \rho_{r,0} 2\frac{A\left(-6(\rho_{d,0}+4/3\rho_{r,0})(2/3\rho_{r,0}^2+(-T/3+7/6\rho_{d,0})\rho_{r,0}-1/4\rho_{d,0} T+1/2\rho_{d,0}^2+A)\sqrt{3}+\mathscr{N}\right)}{(\rho_{d,0}+4/3\rho_{r,0})\left(-6A(T-\rho_{d,0}-2/3\rho_{r,0})\sqrt{3}+\mathscr{N}\right)}},$$ (115)

where

$$\mathscr{J} = 6c\rho_d + 6c\rho_r.$$

We present the behavior of the fractional energy densities in Fig. 6.

### 2.2 Reconstructing modified teleparallel gravity from the GCG model

The frame-work of the GCG in the modified theory of gravity was first proposed by Rastall [40]. Here, we consider the general model of CG model to reconstruct $f(T)$ gravity model and the corresponding thermodynamical quantities of the torsion fluid. Similar to our previous discussions for the OCG model, we consider five cases, namely vacuum, radiation-torsion, dust-torsion, stiff matter-torsion and radiation-dust-torsion systems.

#### 2.2.1 Vacuum case

Here we assume that the vacuum Universe and the energy density and pressure of the matter fluid are negligible, i,e., $p_m = \rho_m = 0$. The general expression of Eq. (17) in vacuum system is given as

$$\left[-\frac{1}{f'}\right]^{1+\alpha} \left[-\frac{1}{2}(f - Tf')\right]^{1+\alpha} = -A,$$ (116)

and we can reconstruct the $f(T)$ gravity model from this expression and it is given by

$$f(T) = c\left[T - 2(-A)^{\frac{1}{1+\alpha}}\right].$$ (117)

In most cases, we have $\alpha$ as a positive constant, with a value between 0 to 1. However, in the literature [41–46], the value of $\alpha$ can be a free parameter and larger than $-1$. If, in our case, $\alpha = 1/2$, the reconstructed $f(T)$ gravity model in Eq. (117) becomes imaginary, $f(T) = cT - 2ic\sqrt{A}$; on the other hand, if $\alpha = -1/2$, then $f(T) = cT - 2c\sqrt{A}$ and this solution is exactly the same as the selected solution in the OCG model, Eq. (23). This is our motivation to account for the parameter $\alpha$ as being either positive or negative. Based on the claim of [41,43–46] and the above motivation, $\alpha$ can be a negative number $\alpha \geq -1$. Then, we choose the parameter $\alpha = -1/2$, the reconstructed $f(T)$ function in GCG in Eq. (117) is reduced to OCG in Eq. (23). Based on this suggestion, we have set the value of $\alpha = -1/2$ in this paper for further investigation of different $f(T)$ gravity models, to manifest the cosmological implication of the corresponding thermodynamical quantities. Therefore, this $f(T)$ which is presented in Eq. (117) is the general of the Eq. (23). By substituting Eq. (117) into Eqs. (7) and (9), it could be reconstruct the energy density, pressure and the corresponding equation of state for torsion fluid in the vacuum system.

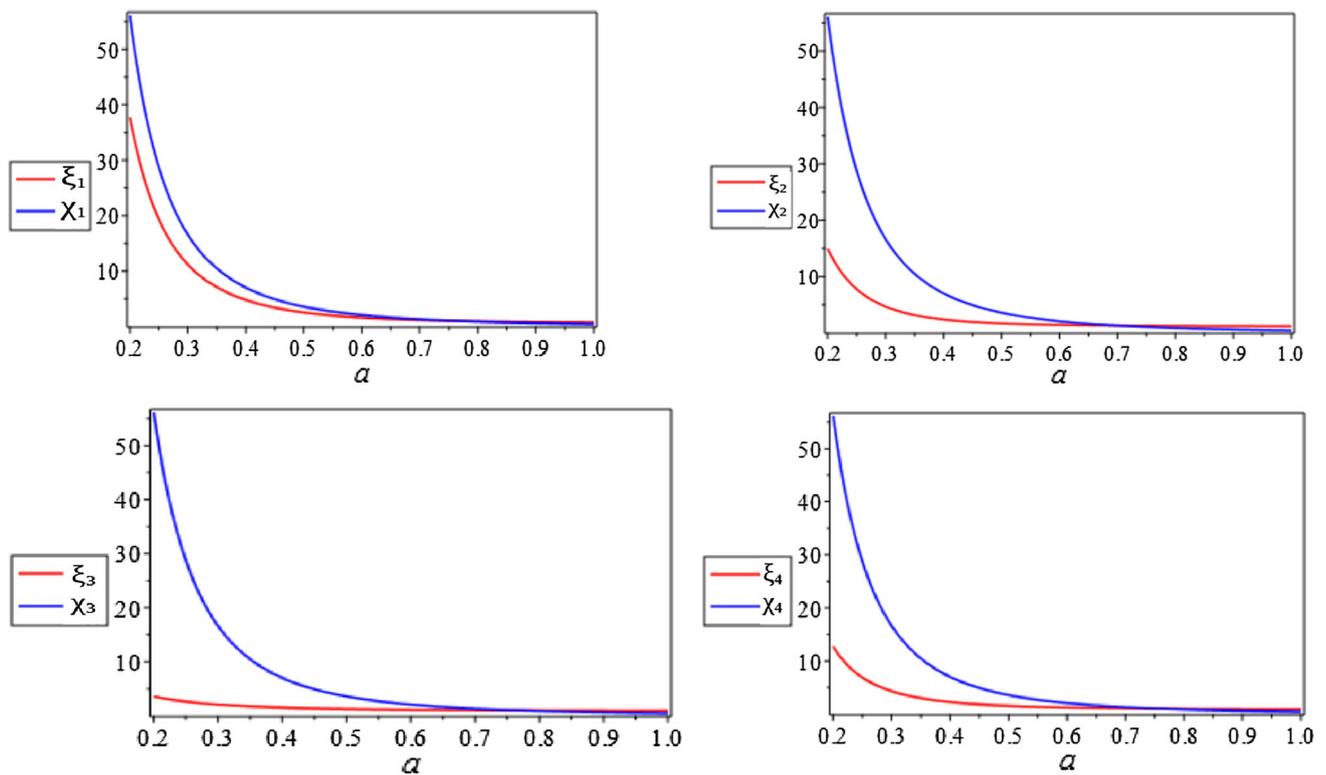

**Fig. 6** All panels show $\xi_{1,2,3,4}$ and $\chi_{1,2,3,4}$ versus $a$ for radiation-dust-torsion system for OCG. We use $c = 0.9$ and $A = 0.2$ for all numerical plotting

### 2.2.2 Radiation-torsion system

Here we consider the radiation-torsion system in GCG model, and radiation with torsion component is considered as an exotic fluid to lead the cosmic expansion. The energy density and pressure of the fluid has been involving from the early to late Universe as a constant and which are presented in Eqs. (7) and (9). The general expression of Eq. (17) for radiation-torsion system is given by

$$-\frac{1}{f'}\left[(f'-1)\frac{\rho_r}{3} - \frac{1}{2}(f-Tf')\right]$$
$$\left[-\frac{1}{f'}\left((f'-1)\rho_r + \frac{1}{2}(f-Tf')\right)\right]^\alpha = -A. \quad (118)$$

Then, by applying the same reasoning as vacuum case, we set the value of $\alpha = -1/2$, to reconstruct different $f(T)$ gravity models in the below:

$$f_1(T) = cT + \frac{1}{3}\left(2c\rho_r + Ac^2\left(3A - \sqrt{3}\sqrt{\frac{3A^2c - 16\rho_r c + 16\rho_r}{c}}\right) - 2\rho_r\right), \quad (119)$$

$$f_2(T) = cT + \frac{1}{3}\left(2c\rho_r + Ac^2\left(3A + \sqrt{3}\sqrt{\frac{3A^2c - 16\rho_r c + 16\rho_r}{c}}\right) - 2\rho_r\right). \quad (120)$$

By substituting $f_1(T)$ and $f_2(T)$ into Eq. (7), we obtain the energy density of the torsion fluid $\rho_T^1$ and $\rho_T^2$. These the energy density of the torsion fluid express as follows:

$$\rho_T^1 = \frac{1}{6c}\left(-A\sqrt{3}\sqrt{\frac{(3A^2 - 16\rho_r)c + 16\rho_r}{c}} + (3A^2 - 8\rho_r)c + 8\rho_r\right), \quad (121)$$

$$\rho_T^2 = \frac{1}{6c}\left(A\sqrt{3}\sqrt{\frac{(3A^2 - 16\rho_r)c + 16\rho_r}{c}} + (3A^2 - 8\rho_r)c + 8\rho_r\right). \quad (122)$$

In the limiting case of $\rho_r = 0$, the function $f_1(T)$ in Eq. (119) and the energy density of the torsion fluid $\rho_1^T$ in Eq. (121) are reduced to vacuum case while, the function $f_2(T)$ in Eq. (120) and the energy density of the torsion fluid $\rho_2^T$ in Eq. (122) go to zero. Here, we also reconstruct the corresponding pressure of the torsion fluid during radiation-torsion system by substituting $f_1(T)$ and $f_2(T)$ into Eq. (9) and we have

$$p_T^1 = \frac{1}{6c}\left(A\sqrt{3}\sqrt{\frac{(3A^2 - 16\rho_r)c + 16\rho_r}{c}} + (-3A^2 - 4\rho_r)c + 4\rho_r\right), \quad (123)$$

$$p_T^2 = \frac{1}{6c}\left(-A\sqrt{3}\sqrt{\frac{(3A^2 - 16\rho_r)c + 16\rho_r}{c}} + (-3A^2 - 4\rho_r)c + 4\rho_r\right), \quad (124)$$

and the equation of state parameter $w_T$ of the torsion fluid is given as

$$w_T^1 = \frac{A\sqrt{3}\sqrt{\frac{(3A^2-16\rho_r)c+16\rho_r}{c}} + (-3A^2 - 4\rho_r)c + 4\rho_r}{-A\sqrt{3}\sqrt{\frac{(3A^2-16\rho_r)c+16\rho_r}{c}} + (3A^2 - 8\rho_r)c + 8\rho_r}, \quad (125)$$

$$w_T^2 = \frac{-A\sqrt{3}\sqrt{\frac{(3A^2-16\rho_r)c+16\rho_r}{c}}c + (-3A^2 - 4\rho_r)c + 4\rho_r}{A\sqrt{3}\sqrt{\frac{(3A^2-16\rho_r)c+16\rho_r}{c}} + (3A^2 - 8\rho_r)c + 8\rho_r}. \quad (126)$$

We use the definition of Eq. (18) and the numerical plots of the equation of state parameter for the radiation-torsion system is presented in Fig. 7 for GCG.

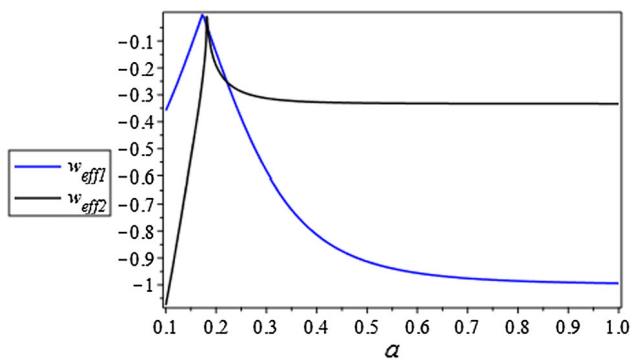

**Fig. 7** $w_{eff(1,2)}$ versus $a$ for radiation-torsion system for GCG. We use $c = 2.5$ and $A = 1$ for all numerical plotting

In the similar mathematical manipulation as OCG in GCG we also obtain the growth factor parameters by taking the ratio of the energy density of torsion fluid $\xi$ and effective fluid $\chi$. These the growth factor parameters are given as follows:

$$\xi_1 = \frac{-A\sqrt{3}\sqrt{\frac{(3A^2-16\rho_r)c+16\rho_r}{c}}c + (3A^2 - 8\rho_r)c + 8\rho_r}{-A\sqrt{3}\sqrt{\frac{(3A^2-16\rho_{r,0})c+16\rho_{r,0}}{c}}c + (3A^2 - 8\rho_{r,0})c + 8\rho_{r,0}}, \quad (127)$$

$$\xi_2 = \frac{A\sqrt{3}\sqrt{\frac{(3A^2-16\rho_r)c+16\rho_r}{c}}c + (3A^2 - 8\rho_r)c + 8\rho_r}{-A\sqrt{3}\sqrt{\frac{(3A^2-16\rho_{r,0})c+16\rho_{r,0}}{c}}c + (3A^2 - 8\rho_{r,0})c + 8\rho_{r,0}}, \quad (128)$$

$$\chi_1 = \frac{\rho_r - A\sqrt{3}\sqrt{\frac{(3A^2-16\rho_r)c+16\rho_r}{c}}c + (3A^2 - 2\rho_r)c + 8\rho_r}{\rho_{r,0} - A\sqrt{3}\sqrt{\frac{(3A^2-16\rho_{r,0})c+16\rho_{r,0}}{c}}c + (3A^2 - 2\rho_{r,0})c + 8\rho_{r,0}}, \quad (129)$$

$$\chi_2 = \frac{\rho_r + A\sqrt{3}\sqrt{\frac{(3A^2-16\rho_r)c+16\rho_r}{c}}c + (3A^2 - 2\rho_r)c + 8\rho_r}{\rho_{r,0} - A\sqrt{3}\sqrt{\frac{(3A^2-16\rho_{r,0})c+16\rho_{r,0}}{c}}c + (3A^2 - 2\rho_{r,0})c + 8\rho_{r,0}}. \quad (130)$$

The Eqs. (127) and (129) are presented on the left side of Fig. 8, and Eqs. (128) and (130) are presented on the right side of Fig. 8.

### 2.2.3 Dust-torsion system

In this section, we also consider that dust-torsion system in GCG model and dust and torsion component are considering as an exotic fluids to reconstruct $f(T)$ gravity model. Then, Eq. (17) is given as

$$[f - Tf']\left[(f' - 1)\rho_d + \frac{1}{2}(f - Tf')\right]^\alpha - 2(-1)^{(1+\alpha)}(f')^{(1+\alpha)}A = 0. \quad (131)$$

By setting the $\alpha = -1/2$, the solutions of Eq. (131) are given as follows:

$$f_1(T) = -c\left(A^2 + A\sqrt{\frac{(A^2 - 4\rho_d)c + 4\rho_d}{c}} - T\right), \quad (132)$$

$$f_2(T) = c\left(A\sqrt{\frac{(A^2 - 4\rho_d)c + 4\rho_d}{c}} - A^2 + T\right). \quad (133)$$

Based on the above reconstructed functions $f_1(T)$ and $f_2(T)$, we also reconstruct the energy density of the torsion fluid as follows

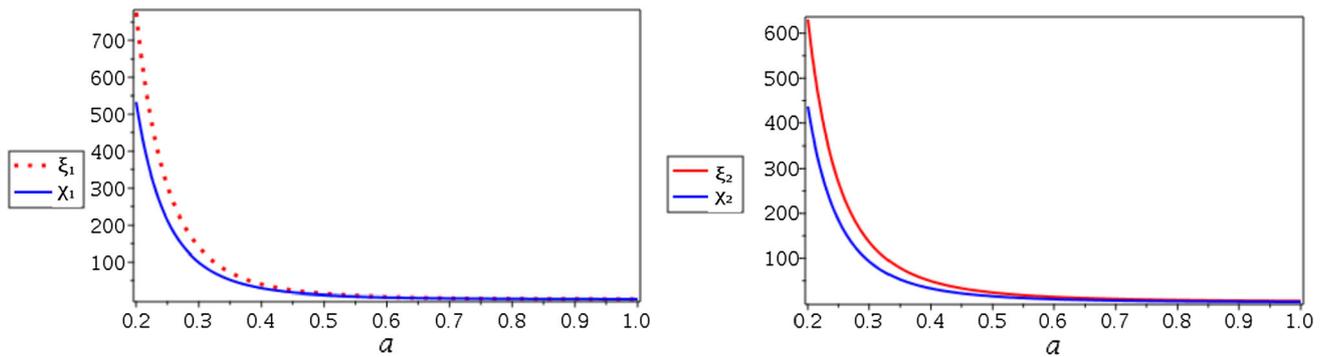

Fig. 8 $\xi_{1,2}$ with $\chi_{1,2}$ versus $a$ for radiation-torsion system for GCG. We use $c = 3$ and $A = 0.8$ for all numerical plotting

$$\rho_T^1 = \frac{1}{2c}\left[(-2\rho_d + A\left(\sqrt{\frac{(A^2 - 4\rho_d)c + 4\rho_d}{c}} + A\right)\right)c + 2\rho_d\right], \tag{134}$$

and

$$\rho_T^2 = \frac{1}{2c}\left[(-2\rho_d - A\left(\sqrt{\frac{(A^2 - 4\rho_d)c + 4\rho_d}{c}} - A\right)\right)c + 2\rho_d\right]. \tag{135}$$

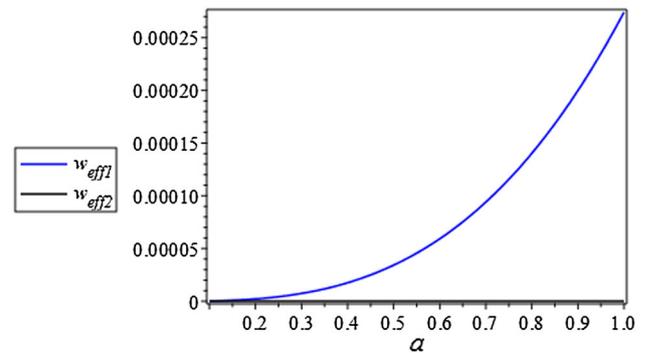

Fig. 9 $w_{eff(1,2)}$ versus $a$ for dust-torsion system for GCG. We use $c = 2$ and $A = 0.98$ for all numerical plotting

By substituting $f_1(T)$ and $f_2(T)$ in Eq. (9) we reconstruct the pressure of the torsion fluid as follows:

$$p_T^1 = \frac{1}{2c}\left[(-2\rho_d - A\left(\sqrt{\frac{(A^2 - 4\rho_d)c + 4\rho_d}{c}} + A\right)\right)c + 2\rho_d\right], \tag{136}$$

$$p_T^2 = \frac{1}{2c}\left[(-2\rho_d + A\left(\sqrt{\frac{(A^2 - 4\rho_d)c + 4\rho_d}{c}} - A\right)\right)c + 2\rho_d\right]. \tag{137}$$

From the reconstructed energy density and pressure of the fluid we have to obtain the following equation of state parameters of torsion fluid in dust dominated case

$$w_T^1 = -\frac{2\rho_d - A\sqrt{\frac{(A^2-4\rho_d)c+4\rho_d}{c}} - Ac - 2\rho_d}{2\rho_d - A\sqrt{\frac{(A^2-4\rho_d)c+4\rho_d}{c}} - Ac - 2\rho_d}, \tag{138}$$

$$w_T^2 = \frac{\left[2\rho_d - A\left(\sqrt{\frac{(A^2-4\rho_d)c+4\rho_d}{c}} - A\right)\right]c - 2\rho_d}{\left[2\rho_d + A\left(\sqrt{\frac{(A^2-4\rho_d)c+4\rho_d}{c}} - A\right)\right]c - 2\rho_d}. \tag{139}$$

We use the definition of Eq. (18) and the numerical plots of the equation of state parameter for the dust-torsion system is presented in Fig. 9 for GCG.

By applying the same reason as radiation dominated case, the growth factor parameters of the fluid are given as follows:

$$\xi_1 = \frac{-A\sqrt{\frac{(A^2-4\rho_d)c+4\rho_d}{c}}c + (A^2 - 2\rho_d)c + 2\rho_d}{\left(A\sqrt{\frac{(A^2-4\rho_{d,0})c+4\rho_{d,0}}{c}} + A^2 - 2\rho_{d,0}\right)c + 2\rho_{d,0}}, \tag{140}$$

$$\xi_2 = \frac{Ac\sqrt{\frac{(A^2-4\rho_d)c+4\rho_d}{c}} + (A^2 - 2\rho_d)c + 2\rho_d}{\left(A\sqrt{\frac{(A^2-4\rho_{d,0})c+4\rho_{d,0}}{c}} + A^2 - 2\rho_{d,0}\right)c + 2\rho_{d,0}}, \tag{141}$$

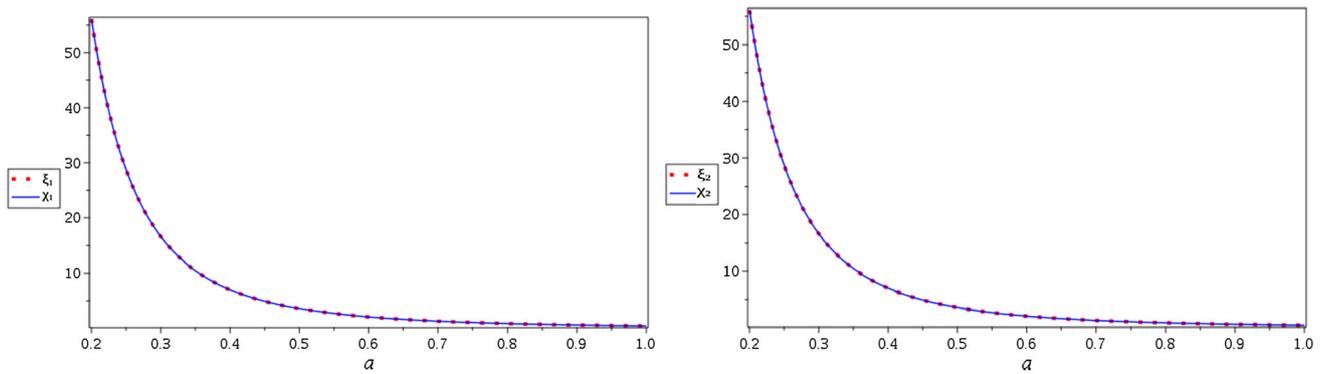

**Fig. 10** All plots show $\xi_{1,2}$ with $\chi_{1,2}$ versus $a$ for dust-torsion system for GCG. We use $c = 3$ and $A = 0.8$ for all numerical plotting

$$\chi_1 = \frac{A^2 c - A\sqrt{\frac{(A^2 - 4\rho_d)c + 4\rho_d}{c}} c + 2\rho_d}{Ac\left(\sqrt{\frac{(A^2 - 4\rho_{d,0})c + 4\rho_{d,0}}{c}} + A\right) + 2\rho_{d,0}}, \quad (142)$$

$$\chi_2 = \frac{A^2 c + Ac\sqrt{\frac{(A^2 - 4\rho_d)c + 4\rho_d}{c}} + 2\rho_d}{Ac\left(\sqrt{\frac{(A^2 - 4\rho_{d,0})c + 4\rho_{d,0}}{c}} + A\right) + 2\rho_{d,0}}. \quad (143)$$

If $\rho_d = 0$, all thermodynamical quantities are reduced to the vacuum case. The Eqs. (140) and (142) are presented on the left side of Fig. 10, and Eqs. (141) and (143) are presented on the right side of Fig. 8.

### 2.2.4 Stiff matter-torsion system

Here, we consider the stiff matter component is considered behind the exotic fluid and the energy density $\rho_m = \rho_s$ has been involving from the early to late Universe as a constant and which are presented in Eqs. (7)–(9). Then, Eq. (17) is given as which gives

$$-\frac{1}{f'}\left[(f' - 1)\rho_s - \frac{1}{2}(f - Tf')\right]$$
$$\left[-\frac{1}{f'}\left((f' - 1)\rho_s + \frac{1}{2}(f - Tf')\right)\right]^\alpha = -A. \quad (144)$$

It can be reconstructed different $f(T)$ gravity models by setting $\alpha = -\frac{1}{2}$ and we have

$$f_1(T) = \left(-\sqrt{\frac{(A^2 - 8\rho_s)c + 8\rho_s}{c}} A\right.$$
$$\left. - A^2 + T + 2\rho_s\right) c - 2\rho_s, \quad (145)$$

$$f(T) = \left(\sqrt{\frac{(A^2 - 8\rho_s)c + 8\rho_s}{c}} A\right.$$
$$\left. - A^2 + T + 2\rho_s\right) c - 2\rho_s, \quad (146)$$

and also by substituting Eqs. (145) and (146) into Eq. (7) we reconstruct the corresponding energy density of torsion. Then we have

$$\rho_T^1 = \frac{1}{2c}\left(\left(-4\rho_s + A\left(\sqrt{\frac{(A^2 - 8\rho_s)c + 8\rho_s}{c}} + A\right)\right) c + 4\rho_s\right), \quad (147)$$

$$\rho_T^2 = \frac{1}{2c}\left(\left(-4\rho_s - A\left(\sqrt{\frac{(A^2 - 8\rho_s)c + 8\rho_s}{c}} - A\right)\right) c + 4\rho_s\right). \quad (148)$$

By substituting Eqs. (145) and (146) into Eq. (9) we can reconstruct the corresponding pressure of torsion. Then we have

$$p_T^1 = -\frac{A}{2}\left(A - \sqrt{\frac{(A^2 - 8\rho_s)c + 8\rho_s}{c}}\right), \quad (149)$$

$$p_T^2 = -\frac{A}{2}\left(A + \sqrt{\frac{(A^2 - 8\rho_s)c + 8\rho_s}{c}}\right), \quad (150)$$

and we also obtain the equation of state parameters as follows:

$$w_T^1 = -\frac{cA\left(A - \sqrt{\frac{(A^2 - 8\rho_s)c + 8\rho_s}{c}}\right)}{\left(\left(-4\rho_s + A\left(\sqrt{\frac{(A^2 - 8\rho_s)c + 8\rho_s}{c}} + A\right)\right)c + 4\rho_s\right)},$$
(151)

$$w_T^2 = \frac{-Ac\sqrt{\frac{(A^2 - 8\rho_s)c + 8\rho_s}{c}} + A}{\left(-4\rho_s + A\left(\sqrt{\frac{(A^2 - 8\rho_s)c + 8\rho_s}{c}} + A\right)\right)c + 4\rho_s}.$$
(152)

### 2.2.5 Radiation-dust-torsion system

In this section we consider the three non interacting fluid, namely dust, radiation and torsion fluid fluid are the cause behind as an exotic fluid for cosmic expansion. The general expression from Eq. (17)

$$-\frac{1}{f'}\left[(f' - 1)p_m - \frac{1}{2}(f - Tf')\right]$$
$$\left[-\frac{1}{f'}\left((f' - 1)\rho_m + \frac{1}{2}(f - Tf')\right)\right]^\alpha = -A.$$
(153)

From this equation we reconstruct two basic $f(T)$ gravity models and these models are given by

$$f_1(T) = Tc + 2cp + A^2$$
$$-\sqrt{\frac{A^2c - 4pc - 4\rho c + 4p + 4\rho}{c}}\bigg) - 2p, \quad (154)$$

$$f_2(T) = Tc + 2cp + A^2$$
$$+\sqrt{\frac{A^2c - 4pc - 4\rho c + 4p + 4\rho}{c}}\bigg) - 2p, \quad (155)$$

where as, based on these constructed $f_1(T)$ and $f_2(T)$ gravity model in Eqs. (154) and (155) we can reconstruct the energy density of the torsion fluid in the non-interacting fluid. By substituting $f_1(T)$ and $f_2(T)$ into Eq. (7), we obtain

$$\rho_T^1 = \frac{1}{2c}\Bigg(\bigg(-2\rho - 2p$$
$$- A\bigg(\sqrt{\frac{(A^2 - 4p - 4\rho)c + 4p + 4\rho}{c}} - A\bigg)\bigg)c + 2p + 2\rho\Bigg),$$
(156)

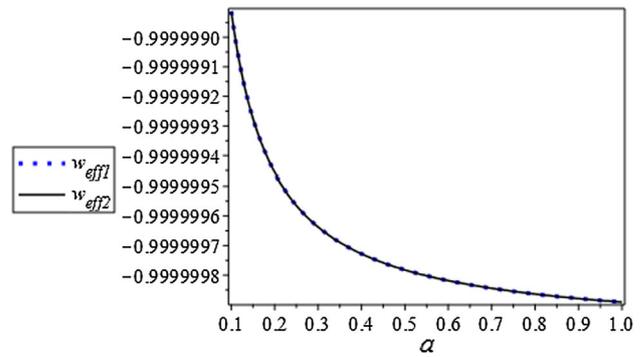

**Fig. 11** $w_{eff(1,2)}$ versus $a$ for radiation- dust-torsion system for GCG. We use $c = 1.3$ and $A = 1$ for all numerical plotting

$$\rho_T^2 = \frac{1}{2c}\Bigg((-2\rho - 2p$$
$$+ A\bigg(\sqrt{\frac{(A^2 - 4p - 4\rho)c + 4p + 4\rho}{c}} + A\bigg)\bigg)c + 2p + 2\rho\Bigg).$$
(157)

Here also we substitute $f_1(T)$ and $f_2(T)$ gravity model in Eqs. (154) and (155) into Eq. (9) to reconstruct the pressure of the torsion fluid in radiation-dust case. Then we have

$$p_T^1 = -\frac{A}{2}\left(A - \sqrt{\frac{(A^2 - 4p - 4\rho)c + 4p + 4\rho}{c}}\right),$$
(158)

$$p_T^2 = -\frac{A}{2}\left(A + \sqrt{\frac{(A^2 - 4p - 4\rho)c + 4p + 4\rho}{c}}\right).$$
(159)

The EoS parameters for torsion are given as follows:

$$w_T^1 = \frac{-Ac\left(\sqrt{\frac{(A^2 - 4p - 4\rho)c + 4p + 4\rho}{c}} - A\right)}{\left[\left(2\rho + 2p + A\left(\sqrt{\frac{(A^2 - 4p - 4\rho)c + 4p + 4\rho}{c}} - A\right)\right)c - 2p - 2\rho\right]},$$
(160)

and

$$w_T^2 = \frac{-Ac\left(\sqrt{\frac{(A^2 - 4p - 4\rho)c + 4p + 4\rho}{c}} + A\right)}{\left[\left(\sqrt{\frac{(A^2 - 4p - 4\rho)c + 4p + 4\rho}{c}}A + A^2 - 2p - 2\rho\right)c + 2p + 2\rho\right]}.$$
(161)

Here, we present the numerical plots of the evolution of effective equation of state parameter in Fig. 11 for radiation-dust-torsion systems.

The effective quantities such as $\xi_{1,2}$ and $\chi_{1,2}$ in the following accordingly:

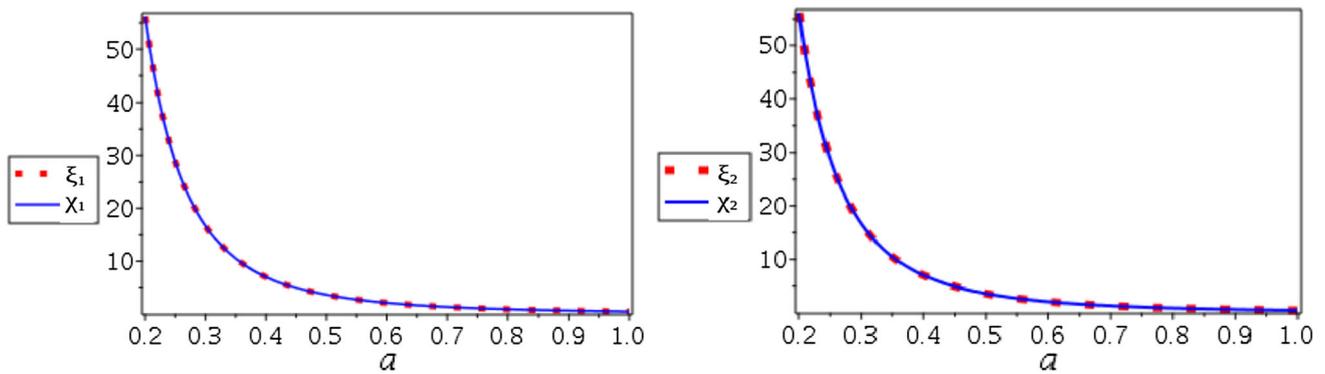

**Fig. 12** In both panels, we present $\xi_{1,2}$ and $\chi_{1,2}$ versus $a$ and use $c = 0.9$ and $A = 0.95$ for these plots

$$\xi_1 = \frac{\left(-cA\sqrt{3}\sqrt{\frac{(3A^2-12\rho_d-16\rho_r)c+12\rho_d+16\rho_r}{c}} + \left(3A^2 - 6\rho_d - 8\rho_r\right)c + 6\rho_d + 8\rho_r\right)}{\left(-cA\sqrt{3}\sqrt{\frac{(3A^2-12\rho_{d,0}-16\rho_{r,0})c+12\rho_{d,0}+16\rho_{r,0}}{c}} + \left(3A^2 - 6\rho_{d,0} - 8\rho_{r,0}\right)c + 6\rho_{d,0} + 8\rho_{r,0}\right)}, \quad (162)$$

$$\xi_2 = \frac{\left(cA\sqrt{3}\sqrt{\frac{(3A^2-12\rho_m-16\rho_r)c+12\rho_m+16\rho_r}{c}} + \left(3A^2 - 6\rho_m - 8\rho_r\right)c + 6\rho_m + 8\rho_r\right)}{\left(-cA\sqrt{3}\sqrt{\frac{(3A^2-12\rho_{d,0}-16\rho_{r,0})c+12\rho_{d,0}+16\rho_{r,0}}{c}} + \left(3A^2 - 6\rho_{d,0} - 8\rho_{r,0}\right)c + 6\rho_{d,0} + 8\rho_{r,0}\right)}, \quad (163)$$

$$\chi_1 = \frac{\left(\rho_d + \rho_r - cA\sqrt{3}\sqrt{\frac{(3A^2-12\rho_d-16\rho_r)c+12\rho_d+16\rho_r}{c}} + \left(3A^2 - 6\rho_d - 8\rho_r\right)c + 6\rho_d + 8\rho_r\right)}{\rho_{d,0} + \rho_{r,0}\left(-cA\sqrt{3}\sqrt{\frac{(3A^2-12\rho_{d,0}-16\rho_{r,0})c+12\rho_{d,0}+16\rho_{r,0}}{c}} + \left(3A^2 - 6\rho_{d,0} - 8\rho_{r,0}\right)c + 6\rho_{d,0} + 8\rho_{r,0}\right)}, \quad (164)$$

$$\chi_2 = \frac{\left(\rho_d + \rho_r cA\sqrt{3}\sqrt{\frac{(3A^2-12\rho_m-16\rho_r)c+12\rho_m+16\rho_r}{c}} + \left(3A^2 - 6\rho_m - 8\rho_r\right)c + 6\rho_m + 8\rho_r\right)}{\rho_{d,0} + \rho_{r,0}\left(-cA\sqrt{3}\sqrt{\frac{(3A^2-12\rho_{d,0}-16\rho_{r,0})c+12\rho_{d,0}+16\rho_{r,0}}{c}} + \left(3A^2 - 6\rho_{d,0} - 8\rho_{r,0}\right)c + 6\rho_{d,0} + 8\rho_{r,0}\right)}. \quad (165)$$

Equations (162) and (164) are presented on the left side of Fig. 12, and Eqs. (163) and (165) are presented on the right side of Fig. 12.

### 2.3 Reconstructing modified teleparallel gravity from the MGCG model

In this section, we consider the generalization of the GCG [27,47–50] in the form:

$$p = \beta\rho - (1+\beta)\frac{A}{\rho^\alpha}, \qquad \beta \neq -1 \, \& \, 0. \quad (166)$$

In analogy with previous sections, the pressure of the torsion fluid in MGCG is given by

$$p_T = \beta\rho_T - (1+\beta)\frac{A}{\rho_T^\alpha}. \quad (167)$$

We substitute $p_T$ and $\rho_T$ from Eqs. (9) and (7) into Eq. (167), obtaining

$$\frac{1}{(-f')^{(1+\alpha)}}\left[(f'-1)p_m - \frac{1}{2}(f - Tf')\right]$$
$$\left[(f'-1)\rho_m + \frac{1}{2}(f - Tf')\right]^\alpha$$
$$= \beta\left[-\frac{1}{f'}\left((f'-1)\rho_m + \frac{1}{2}(f - Tf')\right)\right]^{1+\alpha}$$
$$- (1+\beta)A. \quad (168)$$

Then, this equation is the general expression of Eq. (17); it reduces to Eq. (17) by eliminating the parameter $\beta$. We now reconstruct different $f(T)$ gravity models in the following cases.

### 2.3.1 Vacuum case

In vacuum case, the energy density and pressure of the matter fluid are neglected, $p_m = \rho_m = 0$, and Eq. (168) yields:

$$\frac{1}{(-f')^{(1+\alpha)}} \left[ -\frac{1}{2}(f - Tf') \right] \left[ \frac{1}{2}(f - Tf') \right]^{\alpha}$$
$$= \beta \left[ -\frac{1}{f'} \left( \frac{1}{2}(f - Tf') \right) \right]^{1+\alpha} - (1+\beta)A, \quad (169)$$

the solutions of which are

$$f_1(T) = -2A^2 c + Tc$$
$$f_2(T) = -\frac{c(-\beta^2 + 2\beta - 1)T}{(\beta - 1)^2} - \frac{c(2A^2\beta^2 + 4A^2\beta + 2A^2)}{(\beta - 1)^2}. \quad (170)$$

For $\beta = 0$, the above function $f_2(T)$ is equal $f_1(T)$ and it also coincide with the selected solution of vacuum case in Eq. (23). In order to obtain the energy density of the torsion fluid for this model we substitute Eqs. (170) and (170) into Eq. (7) and it is given by

$$\rho_T^1 = A^2, \qquad \rho_T^2 = \frac{A^2(\beta + 1)^2}{(\beta - 1)^2}, \quad (171)$$

and the corresponding pressures can be found by substitute Eqs. (170) and (170) into Eq. (9) as:

$$p_T^1 = -A^2, \qquad p_T^2 = -\frac{A^2(\beta + 1)^2}{(\beta - 1)^2}. \quad (172)$$

It can be shown that the EoS parameter of the fluid asymptotically approaches that of DE.

### 2.3.2 Radiation-torsion system

In radiation-torsion system in MGCG model, the general expression of the Eq. (168) is

$$\frac{1}{(-f')^{(1+\alpha)}} \left[ (f' - 1)\frac{\rho_r}{3} - \frac{1}{2}(f - Tf') \right]$$
$$\left[ (f' - 1)\rho_r + \frac{1}{2}(f - Tf') \right]^{\alpha}$$
$$= \beta \left[ -\frac{1}{f'} \left( (f' - 1)\rho_r + \frac{1}{2}(f - Tf') \right) \right]^{1+\alpha}$$
$$-(1+\beta)A. \quad (173)$$

By applying the same reasoning as in Sect. 2.2, we set the value of $\alpha = -1/2$ to reconstruct the $f(T)$ gravity models. Then, we obtain

$$f_1(T) = cT - \frac{1}{3(1+\beta)} \Big( 3A^2c\beta + 3A^2c + 6c\beta\rho_r - 2\rho_r c - 6\beta\rho_r + 2\rho_r$$
$$-\sqrt{3}\sqrt{A^2 c(1+\beta)(3A^2c\beta + 3A^2c - 16\rho_r c + 16\rho_r)} \Big), \quad (174)$$

$$f_2(T) = cT - \frac{3A^2c\beta^2 + 6A^2c\beta + 6c\beta^2\rho_r + 3A^2c - 4c\beta\rho_r - 6\beta^2\rho_r - 2\rho_r c + 4\beta\rho_r + 2\rho_r}{3(\beta^2 - 2\beta + 1)}$$
$$+ \frac{\sqrt{3}\sqrt{A^2 c(1+\beta)^2(3A^2c\beta^2 + 6A^2c\beta + 3A^2c + 16c\beta\rho_r - 16\rho_r c - 16\beta\rho_r + 16\rho_r)}}{3(\beta^2 - 2\beta + 1)}. \quad (175)$$

By substituting $f_1(T)$ and $f_2(T)$ gravity models into Eq. (7) we calculate the energy density of the torsion fluid in MGCG as follows:

$$\rho_T^1 = \frac{-3\sqrt{c\left(((1+\beta)A^2 - 16/3\rho_r)c + 16/3\rho_r\right)(1+\beta)A^2} + (3A^2\beta + 3A^2 - 8\rho_r)c + 8\rho_r}{6c(1+\beta)}, \quad (176)$$

$$\rho_T^2 = \frac{-3\sqrt{A^2\left((A^2\beta^2 + (2A^2 + 16/3\rho_r)\beta + A^2 - 16/3\rho_r)c - 16/3\rho_r(\beta - 1)\right)(1+\beta)^2 c}}{6(\beta - 1)^2 c}$$
$$+ \frac{(3A^2\beta^2 + (6A^2 + 8\rho_r)\beta + 3A^2 - 8\rho_r)c - 8\rho_r(\beta - 1)}{6(\beta - 1)^2 c}. \quad (177)$$

In the same manner, by substituting $f_1(T)$ and $f_2(T)$ gravity models into Eq. (9) we calculate the pressure of the torsion fluid in MGCG as follows:

$$p_T^1 = \frac{-12(c-1)(\beta+1/3)\rho_r - 3A^2c\beta - 3A^2c}{6(1+\beta)c}$$
$$+ \frac{3\sqrt{c\left(((1+\beta)A^2 - 16/3\rho_r)c + 16/3\rho_r\right)(1+\beta)A^2}}{6(1+\beta)c}, \quad (178)$$

$$p_T^2 = \frac{-12(c-1)(\beta-1)(\beta-1/3)\rho_r - 3A^2c\beta^2 - 6A^2c\beta - 3A^2c}{6(\beta-1)^2 c}$$
$$+ \frac{3\sqrt{A^2\left((A^2\beta^2 + (2A^2 + 16/3\rho_r)\beta + A^2 - 16/3\rho_r)c - 16/3\rho_r(\beta-1)\right)(1+\beta)^2 c}}{6(\beta-1)^2 c}. \quad (179)$$

From the above energy density and the pressure terms of the fluid we obtain the equation of state parameter of the torsion fluid in MGCG model as follows:

$$w_T^1 = \frac{3\sqrt{c\left(((1+\beta)A^2 - 16/3\rho_r)c + 16/3\rho_r\right)(1+\beta)A^2}}{-3\sqrt{c\left(((1+\beta)A^2 - 16/3\rho_r)c + 16/3\rho_r\right)(1+\beta)A^2} + (3A^2\beta + 3A^2 - 8\rho_r)c + 8\rho_r}$$
$$+ \frac{((-12\beta - 4)\rho_r - 3(1+\beta)A^2)c + (12\beta + 4)\rho_r}{-3\sqrt{c\left(((1+\beta)A^2 - 16/3\rho_r)c + 16/3\rho_r\right)(1+\beta)A^2} + (3A^2\beta + 3A^2 - 8\rho_r)c + 8\rho_r}, \quad (180)$$

$$w_T^2 = \frac{3\sqrt{A^2\left((A^2\beta^2 + (2A^2 + 16/3\rho_r)\beta + A^2 - 16/3\rho_r)c - 16/3\rho_r(\beta-1)\right)(1+\beta)^2 c}}{M_1}$$
$$+ \frac{((-3A^2 - 12\rho_r)\beta^2 + (-6A^2 + 16\rho_r)\beta - 3A^2 - 4\rho_r)c + 12\rho_r(\beta-1)(\beta-1/3)}{M_1}, \quad (181)$$

where

$$M_1 = -3\sqrt{A^2\left((A^2\beta^2 + (2A^2 + 16/3\rho_r)\beta + A^2 - 16/3\rho_r)c - 16/3\rho_r(\beta-1)\right)(1+\beta)^2 c}$$
$$+ \left(3A^2\beta^2 + (6A^2 + 8\rho_r)\beta + 3A^2 - 8\rho_r\right)c - 8\rho_r(\beta-1). \quad (182)$$

The plots in Fig. 13 show the effective equation of state parameter for radiation-torsion system for MGCG model. Then the fractional energy density of the torsion fluid express as follows:

$$\xi_1 = \frac{-3\sqrt{c\left(((1+\beta)A^2 - 16/3\rho_r)c + 16/3\rho_r\right)(1+\beta)A^2} + ((3+3\beta)A^2 - 8\rho_r)c + 8\rho_r}{M_2}, \quad (183)$$

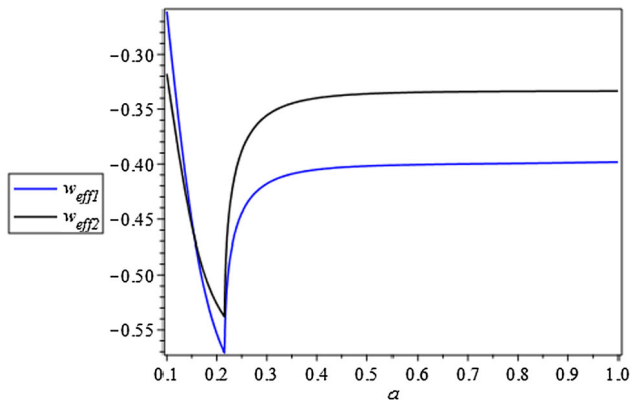

**Fig. 13** $w_{eff(1,2)}$ versus $a$ for radiation-torsion system for MGCG for $\beta = 10$, $c = 1.2$ and $A = 1$

where

$$M_5 = \left(\left(\left(-3A^2 - 6\rho_{r,0}\right)\beta - 3A^2 + 2\rho_{r,0}\right)c - 8\rho_{r,0}\right.$$
$$\left. + 3\sqrt{(1+\beta)cA^2\left(((1+\beta)A^2 - 16/3\rho_{r,0})c + 16/3\rho_{r,0}\right)}\right)(\beta-1)^2.$$

The Eqs. (183) and (185) are presented on the left side of Fig. 14, and Eqs. (184) and (186) are presented on the right side of Fig. 14.

### 2.3.3 Dust-torsion system

As we indicated in the earlier in dust dominated Universe, the energy density of the pressure less fluid acts as a constant behind the exotic fluid from the early to late Universe [46]. In this limit, Eq. (168) will look like

where

$$M_2 = -3\sqrt{A^2 c (1+\beta)\left(((1+\beta)A^2 - 16/3\rho_{r,0})c + 16/3\rho_{r,0}\right)}$$
$$+ \left((3+3\beta)A^2 - 8\rho_{r,0}\right)c + 8\rho_r.$$

$$\xi_2 = \frac{\left(-3A^2 c\beta^2 + \left((-6A^2 - 8\rho_r)c + 8\rho_r\right)\beta + \left(-3A^2 + 8\rho_r\right)c - 8\rho_r\right)(1+\beta)}{M_4}$$
$$+ \frac{3\sqrt{A^2\left(A^2\beta^2 + \left(2A^2 + 16/3\rho_r\right)\beta + A^2 - 16/3\rho_r\right)c - 16/3\rho_r\,(\beta-1)\,(1+\beta)^2 c}}{M_4}, \quad (184)$$

where

$$M_4 = (\beta-1)^2 \left(3\sqrt{(1+\beta)cA^2\left(((1+\beta)A^2 - 16/3\rho_{r,0})c + 16/3\rho_{r,0}\right)}\right.$$
$$\left. - 8\rho_{r,0} - 3A^2 c\beta + \left(-3A^2 + 8\rho_{r,0}\right)c\right),$$

and the ratio of effect energy density of the fluid

$$\chi_1 = \frac{-3\sqrt{c\left(((1+\beta)A^2 - 16/3\rho_r)c + 16/3\rho_r\right)(1+\beta)A^2} + \left((3A^2 + 6\rho_r)\beta + 3A^2 - 2\rho_r\right)c + 8\rho_r}{M_3}, \quad (185)$$

where

$$M_3 = -3\sqrt{A^2 c (1+\beta)\left(((1+\beta)A^2 - 16/3\rho_{r,0})c + 16/3\rho_{r,0}\right)}$$
$$+ \left((3A^2 + 6\rho_{r,0})\beta + 3A^2 - 2\rho_{r,0}\right)c + 8\rho_{r,0},$$

$$\chi_2 = \frac{\left(-3c\left(A^2 + 2\rho_r\right)\beta^2 + \left((-6A^2 + 4\rho_r)c + 8\rho_r\right)\beta + \left(-3A^2 + 2\rho_r\right)c - 8\rho_r\right)(1+\beta)}{M_5}$$
$$+ \frac{3\sqrt{A^2\left(A^2\beta^2 + \left(2A^2 + 16/3\rho_r\right)\beta + A^2 - 16/3\rho_r\right)c - 16/3\rho_r\,(\beta-1)\,(1+\beta)^2 c}}{M_5}, \quad (186)$$

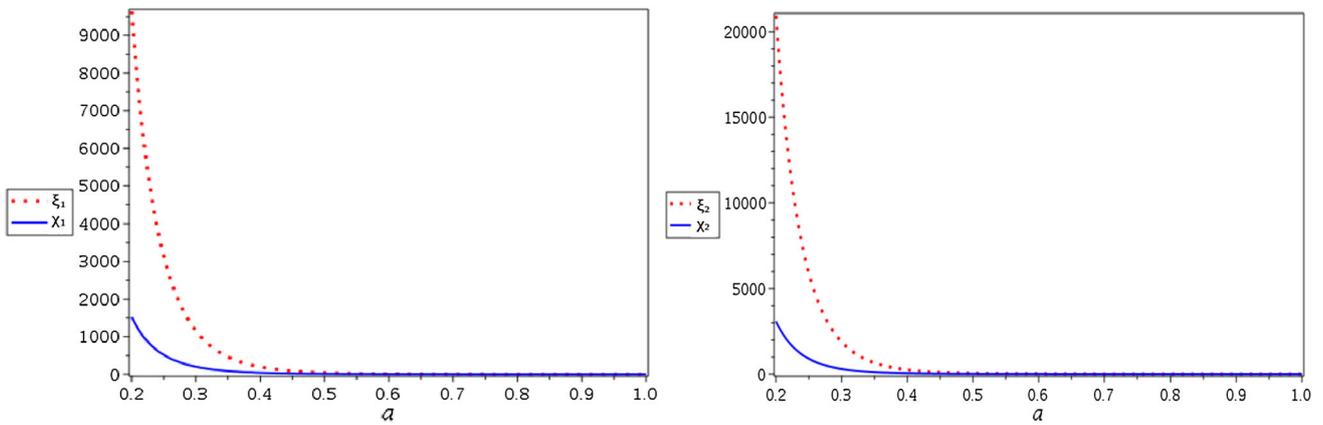

**Fig. 14** All plots shows, $\xi_{1,2,3,4}$ and $\chi_{1,2,3,4}$ versus $a$ for radiation-torsion for MGCG model. We use $\beta = 3$, $c = 0.9$ and $A = 0.8$ for all numerical plotting

$$\frac{1}{(-f')^{(1+\alpha)}} \left[ -\frac{1}{2}(f - Tf') \right] \left[ (f'-1)\rho_d + \frac{1}{2}(f - Tf') \right]^\alpha$$
$$= \beta \left[ -\frac{1}{f'} \left( (f'-1)\rho_d + \frac{1}{2}(f - Tf') \right) \right]^{1+\alpha} - (1+\beta)A, \tag{187}$$

For $\alpha = -\frac{1}{2}$, the solutions as

$$f_1(T) = cT - \frac{1}{1+\beta}\left(A^2 c \beta + A^2 c + 2c\beta\rho_d - 2\beta\rho_d - \sqrt{A^2 c (1+\beta)(A^2 c \beta + A^2 c - 4\rho_d c + 4\rho_d)}\right), \tag{188}$$

$$f_2(T) = cT - \frac{1}{(\beta - 1)^2}\Big[A^2 c \beta^2 + 2A^2 c \beta + 2c\beta^2\rho_d + A^2 c - 2c\beta\rho_d - 2\beta^2\rho_d + 2\beta\rho_d$$
$$- \sqrt{A^2 c (1+\beta)^2 (A^2 c \beta^2 + 2A^2 c \beta + A^2 c + 4c\beta\rho_d - 4\rho_d c - 4\beta\rho_d + 4\rho_d)}\,\Big], \tag{189}$$

and the energy density of the fluid

$$\rho_T^1 = \frac{(-2\rho_d + (1+\beta)A^2)c - \sqrt{(1+\beta)c\left(((1+\beta)A^2 - 4\rho_d)c + 4\rho_d\right)A^2} + 2\rho_d}{2(1+\beta)c}, \tag{190}$$

$$\rho_T^2 = \frac{-\sqrt{c(1+\beta)^2\left((A^2\beta^2 + (2A^2 + 4\rho_d)\beta + A^2 - 4\rho_d)c - 4\rho_d(\beta-1)\right)A^2}}{2(\beta-1)^2 c}$$
$$+ \frac{(A^2\beta^2 + (2A^2 + 2\rho_d)\beta + A^2 - 2\rho_d)c - 2\beta\rho_d + 2\rho_d}{2(\beta-1)^2 c}. \tag{191}$$

And the pressure of the fluid

$$p_T^1 = \frac{\sqrt{(1+\beta)c\left(((1+\beta)A^2 - 4\rho_d)c + 4\rho_d\right)A^2} + \left((-A^2 - 4\rho_d)\beta - A^2 - 2\rho_d\right)c + 2\rho_d(1+2\beta)}{2c(1+\beta)}, \tag{192}$$

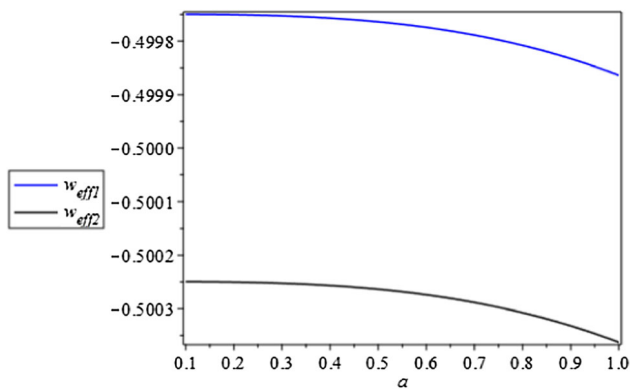

**Fig. 15** $w_{eff(1,2)}$ versus $a$ for dust-torsion system for MGCG for $\beta = 4$, $c = 1.2$ and $A = 1$

where

$$M_8 = \Big[ -\sqrt{(1+\beta)\,c\,A^2\,\{[(1+\beta)A^2 - 16/3\,\rho_{d,0}]c + 16/3\,\rho_{d,0}\}} + \left(A^2\beta + A^2 - 8/3\,\rho_{d,0}\right)c + 8/3\,\rho_{d,0}\Big](\beta-1)^2.$$

The ratio of the effective energy density of the fluid expressed as follows

$$p_T^2 = \frac{-4(\beta-1/2)(\beta-1)(c-1)\rho_d - A^2c\beta^2 - 2A^2c\beta - A^2c}{2(\beta-1)^2 c}$$
$$+ \frac{\sqrt{c(1+\beta)^2\left(\left(A^2\beta^2+(2A^2+4\rho_d)\beta + A^2 - 4\rho_d\right)c - 4\rho_d(\beta-1)\right)A^2}}{2(\beta-1)^2 c}. \tag{193}$$

The corresponding equation of state parameters are give as follows:

$$w_T^1 = \frac{\sqrt{(1+\beta)\,c\,\left(((1+\beta)A^2 - 4\rho_d)c + 4\rho_d\right)A^2} + ((-4\beta-2)\rho_d - A^2\beta - A^2)c + (4\beta+2)\rho_d}{-\sqrt{(1+\beta)\,c\,\left(((1+\beta)A^2 - 4\rho_d)c + 4\rho_d\right)A^2} + (A^2\beta + A^2 - 2\rho_d)c + 2\rho_d}, \tag{194}$$

$$w_T^2 = \frac{4(\beta-1/2)(\beta-1)(c-1)\rho_d + A^2c\beta^2 + 2A^2c\beta + A^2c}{M_6} - \frac{\sqrt{c(1+\beta)^2\left(\left(A^2\beta^2+(2A^2+4\rho_d)\beta + A^2 - 4\rho_d\right)c - 4\rho_d(\beta-1)\right)A^2}}{M_6}, \tag{195}$$

where

$$M_6 = -2(\beta-1)(c-1)\rho_d - A^2c\beta^2 - 2A^2c\beta - A^2c$$
$$+\sqrt{c(1+\beta)^2\left(\left(A^2\beta^2+(2A^2+4\rho_d)\beta + A^2 - 4\rho_d\right)c - 4\rho_d(\beta-1)\right)A^2}$$

Figure 15 shows the effective equation of state parameter for dust-torsion system for MGCG model. The ratio of the energy density in torsion fluid expressed as follows:

$$\xi_1 = \frac{-3\sqrt{(1+\beta)\,c\,\{[(1+\beta)A^2 - 4\rho_d]c + 4\rho_d\}\,A^2} + [(3\beta+3)A^2 - 6\rho_d]c + 6\rho_d}{-3\sqrt{A^2c(1+\beta)\,\{[(1+\beta)A^2 - \tfrac{16}{3}\rho_{d,0}]c + \tfrac{16}{3}\rho_{d,0}\}} + [(3\beta+3)A^2 - 8\rho_{d,0}]c + 8\rho_d}, \tag{196}$$

$$\xi_2 = \frac{\left[-\sqrt{c(1+\beta)^2\left(\left(A^2\beta^2+(2A^2+4\rho_d)\beta + A^2 - 4\rho_d\right)c - 4\rho_d(\beta-1)\right)A^2}\right.}{M_8}$$
$$+\frac{\left.\left(A^2\beta^2+(2A^2+2\rho_d)\beta + A^2 - 2\rho_d\right)c - 2\rho_d(\beta-1)\right](1+\beta)}{M_8}, \tag{197}$$

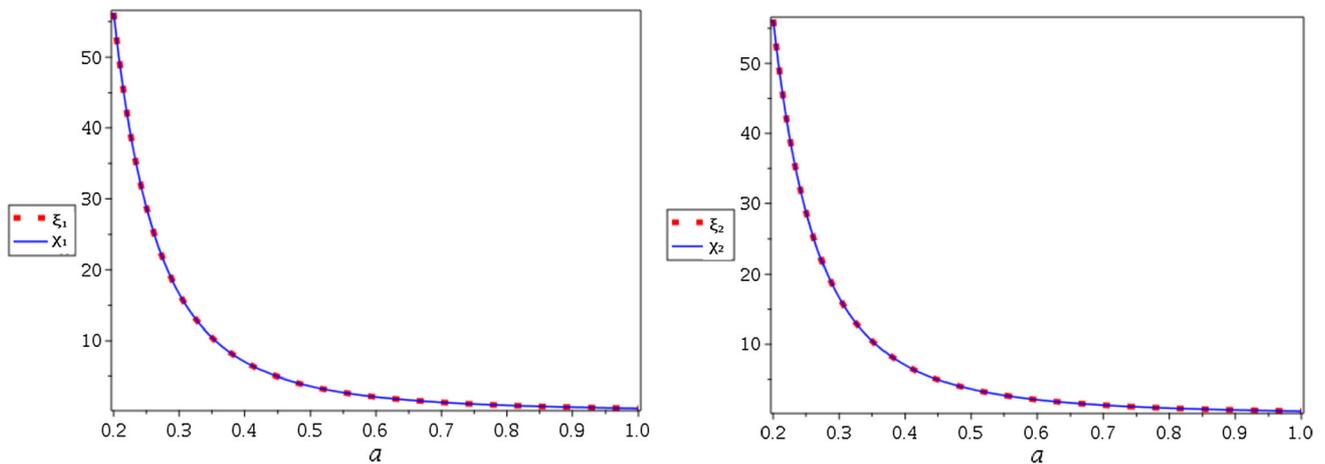

**Fig. 16** $\xi_{1,2,3,4}$ and $\chi_{1,2,3,4}$ versus $a$ for dust-torsion for MGCG model. We use $\beta = 2$, $c = 1.9$ and $A = 0.8$ for all numerical plotting

$$\chi_1 = \frac{-3\sqrt{(1+\beta)c\left\{\left[(1+\beta)A^2 - 4\rho_d\right]c + 4\rho_d\right\}A^2} + \left((3A^2 + 6\rho_d)\beta + 3A^2\right)c + 6\rho_d}{M_7}, \quad (198)$$

where

$$M_7 = -3\sqrt{A^2 c(1+\beta)\left\{\left[(1+\beta)A^2 - 16/3\,\rho_{d,0}\right]c + 16/3\,\rho_{d,0}\right\}} + \left((3A^2 + 6\rho_{d,0})\beta + 3A^2 - 2\rho_{d,0}\right)c + 8\rho_{d,0}$$

$$\chi_2 = \frac{3}{M_9}\Big\{\Big[-\left(A^2 + 2\rho_d\right)c\beta^2 + \left((-2A^2 + 2\rho_d)c + 2\rho_d\right)\beta - A^2 c$$
$$+ \sqrt{c(1+\beta)^2\left((A^2\beta^2 + (2A^2 + 4\rho_d)\beta + A^2 - 4\rho_d)c - 4\rho_d(\beta-1)\right)A^2} - 2\rho_d\Big](1+\beta)\Big\}, \quad (199)$$

where

$$M_9 = (\beta - 1)^2\Big[\left((-3A^2 - 6\rho_{d,0})\beta - 3A^2 + 2\rho_{d,0}\right)c$$
$$+ 3\sqrt{(1+\beta)cA^2\left(((1+\beta)A^2 - 16/3\,\rho_{d,0})c + 16/3\,\rho_{d,0}\right)} - 8\rho_{d,0}\Big].$$

The Eqs. (196) and (198) are presented on the left side of Fig. 16, and Eqs. (197) and (199) are presented on the right side of Fig. 16.

### 2.3.4 Stiff matter-torsion system

Here also we consider the stiff fluid as an exotic fluid in MGCG and the general expression of Eq. (168) and we obtain

$$\frac{1}{(-f')^{(1+\alpha)}}\left[(f'-1)w\rho_s - \frac{1}{2}(f - Tf')\right]$$
$$\left[(f'-1)\rho_d + \frac{1}{2}(f - Tf')\right]^\alpha$$
$$= \beta\left[-\frac{1}{f'}\left((f'-1)\rho_s + \frac{1}{2}(f - Tf')\right)\right]^{1+\alpha}$$
$$-(1+\beta)A, \quad (200)$$

with the solutions of

$$f_1(T) = cT - \frac{A^2 c\beta + A^2 c + 2c\beta\rho_s - 2\rho_s c - 2\beta\rho_s - \sqrt{A^2 c(1+\beta)(A^2 c\beta + A^2 c - 8\rho_s c + 8\rho_s)} + 2\rho_s}{1+\beta} \quad (201)$$

$$f_2(T) = cT - \frac{A^2 c\beta + A^2 c + 2c\beta\rho_s - 2\rho_s c - 2\beta\rho_s}{\beta - 1}$$

$$-\frac{\sqrt{A^2 c(A^2 c\beta^2 + 2A^2 c\beta + A^2 c + 8c\beta\rho_s - 8\rho_s c - 8\beta\rho_s + 8\rho_s)} + 2\rho_s}{\beta - 1}. \quad (202)$$

We reconstruct the energy density by substituting $f_1(T)$ and $f_2(T)$ into Eq. (7) and we obtain

$$\rho_1^T = \frac{((1+\beta)A^2 - 4\rho_s)c - \sqrt{(((1+\beta)A^2 - 8\rho_s)c + 8\rho_s)A^2 c(1+\beta)} + 4\rho_s}{2(1+\beta)c}, \quad (203)$$

$$\rho_T^2 = \frac{A^2 c\beta^2 + ((2A^2 + 4\rho_s)c - 4\rho_s)\beta + (A^2 - 4\rho_s)c + 4\rho_s}{2(\beta-1)^2 c}$$

$$+ \frac{(-\beta - 1)\sqrt{cA^2(c(1+\beta)^2 A^2 + 8\rho_s(\beta - 1)(c - 1))}}{2(\beta-1)^2 c}. \quad (204)$$

Here also we calculate the pressure of the torsion fluid by substituting $f_1(T)$ and $f_2(T)$ into Eq. (9) and we obtain

$$p_T^1 = -\frac{\rho_s(c-1)}{2c(1+\beta)}\Big[A^2 c\beta + A^2 c + 2c\beta\rho_s - 2c\rho_s - 2\beta\rho_s - \sqrt{A^2 c(1+\beta)(A^2 c\beta + A^2 c - 8c\rho_s + 8\rho_s)} + 2c\rho_s\Big],$$

$$p_T^2 = \frac{(1+\beta)\sqrt{c(c(1+\beta)^2 A^2 + 8\rho(\beta-1)(c-1))}A^2 + ((-A^2 - 4\rho)c + 4\rho)\beta^2 + ((-2A^2 + 4\rho)c - 4\rho)\beta - A^2 c}{2(\beta-1)^2 c}.$$

### 2.3.5 Radiation-dust-torsion system

Here is the general expression of radiation-dust-torsion system as an exotic fluid in MGCG model

$$\frac{1}{(-f')^{(1+\alpha)}}\left[(f'-1)p_m - \frac{1}{2}(f - Tf')\right]$$
$$\left[(f'-1)\rho_m + \frac{1}{2}(f - Tf')\right]^\alpha$$
$$= \beta\left[-\frac{1}{f'}\left((f'-1)\rho_m + \frac{1}{2}(f - Tf')\right)\right]^{1+\alpha}$$
$$-(1+\beta)A, \quad (205)$$

with solutions given as

$$f_1(T) = cT - \frac{1}{(\beta-1)^2}\Big[A^2 c\beta^2 + 2A^2 c\beta + 2c\beta^2\rho$$
$$+ A^2 c + 2c\beta p - 2c\beta\rho$$
$$- 2\rho\beta^2 - 2cp - 2\beta p + 2\beta\rho$$
$$- \Big(A^2 c(1+\beta)^2(A^2 c\beta^2 + 2A^2 c\beta + A^2 c$$
$$+ 4c\beta p + 4c\beta\rho - 4cp - 4c\rho - 4\beta p$$
$$- 4\beta\rho + 4p + 4\rho)\Big)^{\frac{1}{2}} + 2p\Big], \quad (206)$$

$$f_2(T) = cT - \frac{1}{1+\beta}\Big[A^2 c\beta + A^2 c$$
$$+ 2c\beta\rho - 2cp - 2\beta\rho$$
$$- \Big(A^2 c(1+\beta)(A^2 c\beta + A^2 c - 4cp$$
$$- 4c\rho + 4p + 4\rho)\Big)^{\frac{1}{2}} + 2p\Big]. \quad (207)$$

From this constructed $f_1(T)$ and $f_2(T)$ gravity model we can reconstruct the energy density of the torsion fluid in the non-interacting fluid. By substituting $f_1(T)$ and $f_2(T)$ into Eq. (7), we obtain

The numerical results of these equation state parameters is presented in Fig. 17 for MGCG model in radiation-dust-torsion system. The the growth factor parameter of energy density of torsion fluid

$$\rho_T^1 = \frac{1}{2(\beta-1)^2 c}\Big[ - \Big(c A^2((A^2\beta^2 + (2A^2 + 4p + 4\rho)\beta + A^2 - 4p - 4\rho)c$$
$$-4(\beta-1)(p+\rho))(1+\beta)^2\Big)^{\frac{1}{2}} + (A^2\beta^2 + (2A^2 + 2p + 2\rho)\beta + A^2 - 2p - 2\rho)c$$
$$+(-2p-2\rho)\beta + 2p + 2\rho\Big] \quad (208)$$

$$\rho_T^2 = \frac{1}{2c(1+\beta)}\Big[ -\sqrt{c\{[(1+\beta)A^2 - 4p - 4\rho]c + 4p + 4\rho\}A^2(1+\beta)}$$
$$+\Big(-2\rho + (1+\beta)A^2 - 2p\Big)c + 2p + 2\rho\Big], \quad (209)$$

and by substituting $f_1(T)$ and $f_2(T)$ into Eq. (9), we obtain pressure of the fluid as follows:

$$p_T^1 = \frac{1}{2(\beta-1)^2 c}\Big[ -A^2 c(1+\beta)^2 - 2\beta(\beta-1)(p+\rho)c + 2\beta(\beta-1)\rho + 2p\beta^2 - 2\beta p$$
$$+\sqrt{c A^2((A^2\beta^2 + (2A^2 + 4p + 4\rho)\beta + A^2 - 4p - 4\rho)c - 4(\beta-1)(p+\rho))(1+\beta)^2}\Big], \quad (210)$$

$$p_T^2 = \frac{1}{2c(1+\beta)}\Big[\sqrt{c\{[(1+\beta)A^2 - 4p - 4\rho]c + 4p + 4\rho\}A^2(1+\beta)}$$
$$+\Big(\Big(-A^2 - 2p - 2\rho\Big)c + 2p + 2\rho\Big)\beta - A^2 c\Big]. \quad (211)$$

The equation of state parameter for torsion fluid become as follows:

$$w_T^1 = \frac{\sqrt{c A^2((A^2\beta^2 + (2A^2 + 4p + 4\rho)\beta + A^2 - 4p - 4\rho)c - 4(\beta-1)(p+\rho))(1+\beta)^2}}{M_{10}}$$
$$+ \frac{((-A^2 - 2p - 2\rho)c + 2p + 2\rho)\beta^2 + ((-2A^2 + 2p + 2\rho)c - 2p - 2\rho)\beta - A^2 c}{M_{10}}, \quad (212)$$

$$w_T^2 = \frac{1}{M_{11}}\Big[\sqrt{c\{[(1+\beta)A^2 - 4p - 4\rho]c + 4p + 4\rho\}A^2(1+\beta)}$$
$$+\Big(\Big(-A^2 - 2p - 2\rho\Big)\beta - A^2\Big)c + (2\rho + 2p)\beta\Big],$$

where $M_{10} = A^2 c\beta^2 + ((2A^2 + 2p + 2\rho)c - 2p - 2\rho)\beta + (A^2 - 2p - 2\rho)c + 2p + 2\rho$

$-\sqrt{c A^2\{[A^2\beta^2 + (2A^2 + 4p + 4\rho)\beta + A^2 - 4p - 4\rho]c - 4(\beta-1)(p+\rho]\}(1+\beta)^2}$

and $M_{11} = \Big(A^2\beta + A^2 - 2p - 2\rho\Big)c + 2p + 2\rho - \sqrt{c\{[(1+\beta)A^2 - 4p - 4\rho]c + 4p + 4\rho\}A^2(1+\beta)}$.

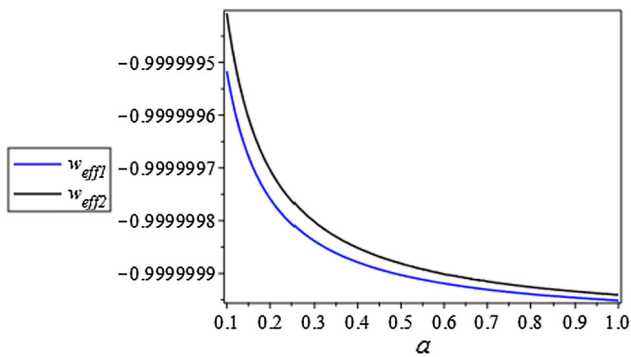

**Fig. 17** $w_{eff(1,2)}$ versus $a$ for radiation-dust-torsion system for MGCG for $\beta = 4$, $c = 1.2$ and $A = 1$

where $M_{15} = -3$
$\sqrt{(1+\beta)c A^2 \{[(1+\beta)A^2 - 16/3\rho_0]c + 16/3\rho_0\}}$
$+ ((3A^2 + 6\rho_0)\beta + 3A^2 - 2\rho_0)c + 8\rho_0$. The Eqs. (213) and (215) are presented on the left side of Fig. 18, and Eqs. (214) and (216) are presented on the right side of Fig. 18.

## 3 Reconstructing $f(T)$ gravity models: the special case

In Sect. 2, we reconstructed Lagrangian densities of modified teleparallel gravity models which depend on both the torsion and matter fluids that resulted in the general

$$\xi_1 = \frac{(1+\beta)}{M_{12}}\Big[\Big(A^2\beta^2 + \big(2A^2 + 2p + 2\rho\big)\beta + A^2 - 2p - 2\rho\Big)c - 2(\beta-1)(p+\rho) \\ - \sqrt{c A^2\{[A^2\beta^2 + (2A^2 + 4p + 4\rho)\beta + A^2 - 4p - 4\rho]c - 4(\beta-1)(p+\rho)\}(1+\beta)^2}\Big], \quad (213)$$

where $M_{12} = (\beta-1)^2\Big[-\sqrt{(1+\beta)c A^2(((1+\beta)A^2 - 16/3\rho_0)c + 16/3\rho_0)} + \big(A^2\beta + A^2 - 8/3\rho_0\big)c + 8/3\rho_0\Big]$

and

$$\xi_2 = \frac{1}{M_{14}}\Big[-3\sqrt{c\{[(1+\beta)A^2 - 4p - 4\rho]c + 4p + 4\rho\}A^2(1+\beta)} + ((3\beta + 3)A^2 - 6p - 6\rho)c + 6p + 6\rho\Big], \quad (214)$$

where $M_{14} = -3$
$\sqrt{(1+\beta)c A^2\{[(1+\beta)A^2 - 16/3\rho_0]c + 16/3\rho_0\}} + ((3\beta + 3)A^2 - 8\rho_0)c + 8\rho_0$. And the growth factor parameters for effective fluid become:

expression (17). Here, we reconstruct some Lagrangian densities of $f(T)$ gravity models with only torsion dependence based on some assumption applied to the modified Raychaudhuri equation (6)

$$\chi_1 = \frac{3}{M_{13}}\Big[\Big(-\big(A^2 + 2\rho\big)c\beta^2 + \big((-2A^2 - 2p + 2\rho)c + 2p + 2\rho\big)\beta + \big(-A^2 + 2p\big)c \\ + \sqrt{c A^2\{[A^2\beta^2 + (2A^2 + 4p + 4\rho)\beta + A^2 - 4p - 4\rho]c - 4(\beta-1)(p+\rho)\}(1+\beta)^2} \\ - 2p - 2\rho\Big)(1+\beta)\Big], \quad (215)$$

where $M_{13} = (\beta-1)^2\Big[\big((-3A^2 - 6\rho_0)\beta - 3A^2 + 2\rho_0\big) c + 3\sqrt{(1+\beta)c A^2(((1+\beta)A^2 - 16/3\rho_0)c + 16/3\rho_0)} - 8\rho_0\Big]$, and

$$\frac{\dot T}{6H} + 3H^2 = \frac{p_m}{f'} + \frac{1}{2f'}(f - Tf') + \frac{4f'' H \dot T}{f'}. \quad (217)$$

$$\chi_2 = \frac{1}{M_{15}}\Big[-3\sqrt{c\{[(1+\beta)A^2 - 4p - 4\rho]c + 4p + 4\rho\}A^2(1+\beta)} \\ + \Big(\big(3A^2 + 6\rho\big)\beta + 3A^2 - 6p\Big)c + 6p + 6\rho\Big], \quad (216)$$

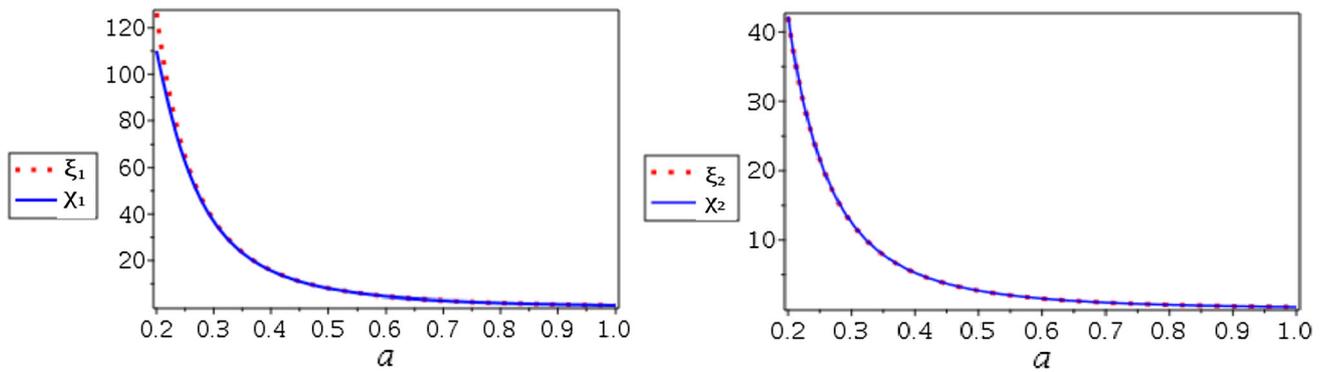

**Fig. 18** $\xi_{1,2,3,4}$ and $\chi_{1,2,3,4}$ versus $a$ for radiation-dust-torsion for MGCG model. We use $\beta = 10$, $c = 1.9$ and $A = 0.8$ for these plots

Assuming as in the previous case $\dot{T} \approx 0$, we can re-write the Friedmann equation as

$$3H^2 \approx \frac{w\rho_m}{f'} + \frac{1}{2f'}(f - Tf'). \tag{218}$$

By substituting Eq. (5) into Eq. (218), we obtain the matter energy density as

$$\rho_m = \frac{f - Tf'}{1 - w}, \quad w \neq 1. \tag{219}$$

Here we observe that, the above definition cannot accommodate the stiff fluid.[5] Then, the corresponding energy density of the torsion fluid reads

$$\rho_T = -\frac{f - Tf'}{f'}\left[\frac{2f' - 1 - w}{2(1 - w)}\right], \tag{220}$$

and the pressure of the torsion fluid as

$$p_T = -\frac{f - Tf'}{f'}\left[\frac{2wf' - w - 1}{2(1 - w)}\right]. \tag{221}$$

From the characteristic equation of state for CG and the expression of the thermodynamic quantities Eqs. (220) and (221), we have

$$\left[\frac{f - Tf'}{f'}\left(\frac{2wf' - w - 1}{2(1 - w)}\right)\right]\left[\frac{f - Tf'}{f'}\left(\frac{2f' - 1 - w}{2(1 - w)}\right)\right]^\alpha = -A. \tag{222}$$

We present the reconstructed different Lagrangian density of $f(T)$ gravity models from Eq. (222) for OCG and GCG models as we did in the previous sections. Unfortunately, the result of MGCG model from this equation seems non-physical solutions. Because of this, we only present the results of OCG and GCG in the subsections.

---
[5] The results for vacuum case is the same as with the previous sections.

### 3.1 For OCG model

The reconstructed Lagrangian density of $f(T)$ gravity models from Eq. (222) for radiation-torsion system are given as:

$$f_1(T) = cT + \frac{2\sqrt{4Ac - 3Ac^2}}{4Ac - 3Ac^2}, \tag{223}$$

$$f_2(T) = -cT - \frac{2\sqrt{4Ac - 3Ac^2}}{4Ac - 3Ac^2}, \tag{224}$$

$$f_3(T) = RootOf\Big(27Z^4 - 108Z^3T + (144T^2 + 72A)Z^2 + (240AT - 64T^3)Z - 16AT^2 + 48A^2\Big). \tag{225}$$

and for dust-torsion system reads as

$$f_1(T) = cT + \frac{2\sqrt{3AC - 2Ac}}{2c - 3}, \tag{226}$$

$$f_2(T) = -cT - \frac{2\sqrt{3AC - 2Ac}}{2c - 3}, \tag{227}$$

$$f_3(T) = \frac{1}{6}(3T^2 - A) + \frac{1}{9}\Bigg[A\Big(729\,T^4 + 270\,AT^2 - 2\,A^2 + 3\sqrt{3}\sqrt{(27\,T^2 - 4\,A)^3\,T}\Big)\Bigg]^{\frac{1}{3}}. \tag{228}$$

### 3.2 For GCG model

Here, we also reconstruct the Lagrangian density of $f(T)$ gravity for radiation-torsion

$$f(T) = cT - A^2\frac{6c - 8}{c}, \quad \text{and} \tag{229}$$

for dust-torsion system as

$$f(T) = cT + 2cA^2. \tag{230}$$

for the GCG model. Therefore, we show another possibility to reconstruct Lagrangian density of $f(T)$ gravity models which depends only on the torsion scalar by defining the energy density of the matter fluid $\rho_m$ as Eq. (219).

## 4 Discussion and conclusions

In this manuscript, we have looked at a Chaplygin-gas-inspired reconstruction of $f(T)$ gravity Lagrangians and their possible implications to cosmological evolution. We considered the most common toy models of the cosmological CG: original, generalized and modified generalized. For each toy model, five cases are taken into account, namely the vacuum, radiation, dust, stiff and radiation-dust systems. We report that for the vacuum case, with the original CG model, the $f(T)$ Lagrangian has the form similar to that of GR. For radiation-torsion and dust-torsion systems, the $f(T)$ Lagrangian possesses a degeneracy of solutions. The system composed of radiation, dust and torsion all together has four $f(T)$ solutions. However, the effect of the mixture cannot easily be seen directly from the solutions. But for the first two $f(T)$ solutions, one has the linear behavior of the torsion T. For the other two solutions, the presence of quadratic terms is manifested.

The consideration of stiff fluid together with torsion brings $f(T)$ Lagrangians where the first two $f(T)$ solutions are linear but the next two possess quadratic terms together with stiff density at the denominator. The implication for this behavior is that once the stiff fluid is small, the two solutions grow in opposite directions based on the presence of a negative sign on the last solution.

The consideration of the generalized CG brings the $f(T)$ Lagrangian with a couple of features. For instance, the vacuum universe has the $f(T)$ Lagrangian with a dependence on $\alpha$, a parameter that constrains the behavior of the cosmological constant if one is interested in the reduction of this Lagrangian to that of GR. Both radiation-torsion and dust-torsion systems possess $f(T)$ Lagrangians spanned by two solutions and there is no presence of the quadratic terms in those solutions. However, using these Lagrangians, the obtained $\rho_T$ and $p_T$ are independent of the torsion $T$. The stiff torsion-fluid system also has two $f(T)$ solutions with features that are similar to those of the radiation and dust systems. The consideration of the system that has all components (radiation, dust, and torsion), has $f(T)$ solutions that have a dependence on the total pressure of standard matter.

For the third type of the CG, the modified generalized one, the vacuum case is considered where the $f(T)$ solution comes with no presence of quadratic terms in $T$. Both radiation and dust systems have two $f(T)$ two solutions, and the same story is manifested for both stiff and the system that has radiation, dust and torsion as components. However, when the consideration of both generalized Chaplygin and modified Chaplygin models is done, we followed the suggestion presented in the literature that the value of $\alpha$ has to be negative in favour of the observations. We consequently selected $\alpha = -\frac{1}{2}$ for our purpose. The choice of other values of $\alpha$ in the suggested range is needed to constrain the $f(T)$ Lagrangians.

We obtained the expressions for the equation of state parameter for the considered $f(T)$ solutions and the corresponding effective equation of state parameters for all cases in each CG model and plotted the results. From the plots, we observed that our reconstructed $f(T)$ gravity models replacing exotic Chaplygin-gas (CG) fluid models can, in principle, explain the currently observed cosmic evolutionary dynamics. For instance, in Fig. 1 for radiation-torsion for OCG, in Fig. 5 for radiation-dust-torsion system for OCG, in Fig. 7 a radiation-torsion system for GCG models (in the limits of $1 \geq a > 0.2$), in Fig. 11 for a radiation-dust-torsion system for GCG, in Fig. 13 for a radiation-torsion system for MGCG, in Fig. 15 for a dust-torsion system for GCG and in (17) for a radiation-dust-torsion system for MGCG models, the evolution equation of state parameters is less than negative one-third ($w_{eff(1,2)} < -\frac{1}{3}$) and asymptotically behaves like that of the equation of state parameter of the cosmological constant fluid $w = -1$. This indicates that torsion fluid replacing the Chaplygin fluid can explain the accelerating expansion of the Universe without cosmological constant scenario. On the other hand, $w_{eff(1,2,3,4)}$ in Fig. 3 for a dust-torsion system for OCG, in Fig. 9 for a dust-torsion system for GCG, $w_{eff(3,4)}$ in Fig. 1 for a radiation-torsion system for OCG, in Fig. 5 for a radiation-dust-torsion system for OCG, in Fig. 9 for dust-torsion system for GCG, the effective evolution equation of state parameters found in the intervals of $-\frac{1}{3} < w_{eff}$ show that the Universe undergoes a decelerated cosmic expansion such as is the case in the early expansion phases of a Chaplygin-gas-dominated universe.

A key assumption made in this work in order to obtain the exact solutions presented is that the torsion scalar T changes slowly over time. The general case of a time-dependent torsion requires more rigorous computational solving techniques, with carefully selected initial conditions. This task and the constraining of the free parameters appearing in our models with existing and future observational data is left for a forthcoming investigation.

**Acknowledgements** SS gratefully acknowledges financial support from Wolkite University, Entoto Observatory and Research Center and Ethiopian Space Science and Technology Institute, as well as the hospitality of the Physics Department of North-West University (NWU) during the preparation of part of this manuscript. JN gratefully acknowledges financial support from the Swedish International Development Cooperation Agency (SIDA) through the International Science Program (ISP) to the University of Rwanda (Rwanda Astrophysics, Space and Climate Science Research Group), project number RWA01. ME acknowledges that this work is supported by an NWU/NRF postdoctoral fellowship. AA acknowledges that this work is based on the research supported in part by the National Research Foundation (NRF) of South Africa with Grant number 112131. SS, JN and AA are grateful for the Institute of Theoretical Astrophysics, University of Oslo, for hosting them during the conceptualization stage of this project.

**Data Availability Statement** This manuscript has no associated data or the data will not be deposited. [Authors' comment: Because the study focused on the theoretical works.]